\documentclass[11pt]{article}
\usepackage[margin=1in]{geometry}
\usepackage[english]{babel}
\usepackage{url}
\usepackage{graphicx}
\usepackage[export]{adjustbox}
\usepackage{blindtext}
\usepackage{caption}
\usepackage{algpseudocode}
\usepackage{algorithm}
\usepackage{float}
\usepackage{mathtools}
\usepackage{amssymb}
\usepackage{amsthm}
\usepackage[dvipsnames]{xcolor}
\usepackage{comment}
\usepackage{multirow}
\usepackage{color, colortbl}
\usepackage{cite}
\usepackage{amsmath}
\theoremstyle{plain}
\newtheorem{theorem}{Theorem}[section]
\newtheorem{proposition}[theorem]{Proposition}
\newtheorem{lemma}[theorem]{Lemma}
\newtheorem{corollary}[theorem]{Corollary}
\theoremstyle{definition}
\newtheorem{definition}[theorem]{Definition}
\newtheorem{assumption}[theorem]{Assumption}
\theoremstyle{remark}
\newtheorem{remark}[theorem]{Remark}

\usepackage{cancel}
\usepackage{subcaption}
\usepackage{booktabs}

\usepackage{mathrsfs}
\newcommand{\keywords}[1]{%
	\par\medskip
	\noindent\textbf{Keywords: }#1
}

\title{Adjoint Kernel Sensitivity of the\\
Rough Heston Characteristic Exponent}
\author{Carlos Escudero \and Francisco J. Luguera-Parúas \and Miguel Sama}
\date{\today}

\begin{document}
	
	\maketitle
	

\begin{abstract}
Multi-factor approximations of rough volatility kernels affect the characteristic exponents used in Fourier-based derivative valuation. We derive an adjoint representation for the first-order sensitivity of the rough Heston characteristic exponent with respect to its memory kernel, with model parameters and the input variance curve held fixed. The Volterra--Riccati functional $\mathcal L_T(K,z)$ admits a local real-analytic extension, jointly in the kernel perturbation and the transform parameter. The Fr\'echet derivative of this extension with respect to the kernel in $L^1$ is represented by integration against a continuous sensitivity kernel $B_K(u,z)$, obtained from a backward linear Volterra equation and independent of the perturbation direction. This representation resolves the leading-order contributions of kernel perturbations across memory lags. On the structural kernel class, which contains the fractional kernel and its multi-factor approximations and has empty interior in $L^1$, the extension yields Hadamard directional differentiability and a first-order expansion with a quadratic $L^1$-remainder for all sufficiently close structural kernels, uniformly on compact subsets of the strip of transform parameters. The same sensitivity kernel yields multi-factor weight and decay-rate derivatives. Numerical experiments at multi-factor reference kernels assess the finite-resolution implementation and illustrate how the lag distribution of positive perturbations with equal $L^1$ mass affects the first-order response.
\end{abstract}

\keywords{rough Heston model; kernel sensitivity;
    adjoint sensitivity analysis; Volterra--Riccati equations;
    Fr\'echet differentiability; multi-factor approximation.}

\par\medskip
\noindent\textbf{2020 Mathematics Subject Classification:}
91G30, 45D05, 46G05, 91G60.

\section{Introduction}
\label{sec:introduction}

Stochastic volatility models describe the random evolution of asset price volatility and provide a framework for derivative valuation and risk management. In models with memory, a kernel governs how past fluctuations influence current volatility, making the effect of kernel perturbations relevant both to model sensitivity and to numerical approximation.
Rough volatility models reproduce the empirical roughness of realized variance 
and the steep short-maturity implied-volatility skew through singular kernels in the volatility dynamics \cite{gatheral_jaisson_rosenbaum_2018, bayer_friz_gatheral_roughvol_2016}. In 
the rough Heston model of El~Euch and Rosenbaum \cite{characteristic_RH}, the 
Fourier--Laplace transform of the log-price is exponentially affine, and its 
logarithm is an integral functional of the solution of a Volterra--Riccati 
equation driven by the memory kernel $K$. In general, no closed-form solution is known for this equation and the fractional kernel is singular at the origin; common numerical 
approaches either discretize the fractional Riccati equation directly 
\cite{CallegaroGrasselliPages2020} or replace $K$ by a more tractable kernel, 
typically a positive sum of exponentials \cite{abi_jaber_el_euch_2019b,abi_jaber_lifting_2019}. This raises the question studied here: how does the characteristic
exponent respond, to first order, to a perturbation of the kernel, and how
is this response distributed across memory lags?

Existing results address this question mainly through norm-based error bounds. 
For rough Heston, Abi~Jaber and El~Euch \cite{abi_jaber_el_euch_2019b} bound 
the deviation of the Riccati solution and, for a modified multi-factor 
approximation, of call prices by the $L^1$-distance between the kernels, while 
Bayer and Breneis \cite{bayer_breneis} obtain $L^1$-based weak error bounds 
for broad classes of European payoffs. Strong error bounds in terms of the 
$L^2$-distance between kernels are available for stochastic Volterra equations with Lipschitz  coefficients \cite{AlfonsiKebaier2024,BayerBreneis2023}, an assumption that 
excludes the square-root coefficient of rough Heston. For a class of 
integrated Volterra processes, Bossy, Mart\'inez and Maurer 
\cite{bossi_maurer} derive a weak kernel-comparison bound through a path-dependent PDE. These results control the size of the error through a norm of the kernel difference; they are not designed to identify its leading-order term, nor how that term depends on where the two kernels differ.

Related sensitivity analyses consider different perturbation variables. Hager and Kreher 
\cite{hager_kreher} prove analyticity of the fractional Riccati solution in 
the Hurst parameter $H$ and expand it along the one-parameter family of fractional kernels
$H\mapsto K_H$; in their parametrization the input curve also varies with $H$. Friz and Gatheral~\cite{friz_gatheral_ssr_2025} use functional derivatives with respect to the forward variance curve, not the memory kernel, in a characteristic-function representation of the skew-stickiness ratio. Pannier \cite{pannier_2026} derives integration-by-parts representations for sensitivities with respect to the initial curve of stochastic Volterra processes, with applications to forward variance curve sensitivities. The robustness and sensitivity analysis of Matas and Posp\'i\v{s}il \cite{matas_pospisil_robustness} is of a different, statistical nature, concerning the variability of calibrated prices and parameters of rough Volterra models under changes in the option data. We are not aware of a result deriving the first variation of the rough Heston characteristic exponent with respect to the kernel itself, in general $L^1$-directions rather than along a parametric family and with an explicit 
representing density; this is the gap addressed here.

We study this first variation in the Volterra--Heston formulation of 
\cite{markovian_rough_volatility}, in which the input variance curve $g$ is a 
datum of the model. For a fixed maturity $T>0$, we consider complex transform parameters
in the strip $\mathcal Z:=\{z\in\mathbb C:\Re z\in[0,1]\}$. Throughout, the mean-reversion coefficient $\lambda$, the volatility-of-volatility parameter $\nu$, the correlation
parameter $\rho$ and the curve $g$ are held fixed, and only
the kernel is perturbed, so that the kernel acts on the functional
\begin{equation*}
	\mathcal L_T(K,z)
	=
	\int_0^T F\bigl(z,\psi_K(T-r,z)\bigr)\,g(r)\,dr,
	\qquad z\in\mathcal Z,
\end{equation*}
only through the Riccati solution $\psi_K$; here $F$ is the quadratic 
polynomial \eqref{eq:F_def} and $K$ ranges over a structural class $\mathscr 
K_T$ (Definition~\ref{def:structural_class})
containing the fractional kernel and its positive multi-factor 
approximations. This convention differs from holding fixed
the initial variance and mean-reversion level of the classical 
parametrization, in which $g$ is recomputed from the kernel 
(Remark~\ref{rem:classical_form}), as in \cite{abi_jaber_el_euch_2019b, 
hager_kreher}; it also differs from holding the forward variance curve $t\mapsto\mathbb E[V_t]$ fixed,
where $V$ denotes the variance process; see \cite{gatheral_keller_ressel_2019} for the affine forward variance framework. The analysis is deterministic: the interpretation of $\exp\mathcal L_T(K,z)$ as
the transform of the log-return $\log(S_T/S_0)$, where $S$
denotes the discounted asset price, requires additional conditions
(Section~\ref{subsec:probabilistic_interpretation}) and is not asserted for 
general perturbed kernels.

Our main result is an adjoint representation of the first variation
with respect to the memory kernel. For a reference kernel $K$
satisfying Assumption~\ref{ass:reference_riccati},
Theorem~\ref{thm:frechet_diff} identifies the Fr\'echet derivative
of the extended Riccati functional with the bounded real-linear
functional
$h\mapsto\int_0^T B_K(u,z)h(u)\,du$ on $L^1_{\mathbb R}(0,T)$.
Here $h$ is a real-valued integrable kernel perturbation and $u$
denotes the memory lag.
Following the adjoint sensitivity principle
\cite{giles_pierce_2000}, the continuous sensitivity kernel
$B_K(\cdot,z)$ is obtained from one forward Riccati solve and
one backward linear Volterra solve for each $z\in\mathcal Z$.
The same kernel evaluates different perturbation directions
and yields the multi-factor weight and decay-rate derivatives
(Remark~\ref{rem:multifactor_sensitivities}).

The construction relies on a joint local real-analytic extension
in the kernel perturbation and the transform parameter
(Theorem~\ref{thm:local_analytic_solution} and
Proposition~\ref{prop:extended_functional_analytic}).
Every structural reference kernel satisfies the required
solvability assumption. Restricting this extension to $\mathscr K_T$
yields Hadamard directional differentiability along its contingent
directions, that is, limits in $L^1$ of directions $h_n$ for which
$K+t_n h_n\in\mathscr K_T$ with $t_n\downarrow0$,
with the same representing functional
(Corollary~\ref{090926}).
The analytic extension also gives a uniform quadratic remainder:
for every $K\in\mathscr K_T$ and every compact set
$\mathfrak K\subset\mathcal Z$, there exist constants
$r_{\mathcal L,\mathfrak K},C_{\mathcal L,\mathfrak K}>0$
such that
\begin{equation*}
    \sup_{z\in\mathfrak K}
    \left|
        \mathcal L_T(K+\delta K,z)-\mathcal L_T(K,z)
        -\int_0^T B_K(u,z)\,\delta K(u)\,du
    \right|
    \le
    C_{\mathcal L,\mathfrak K}
    \|\delta K\|_{L^1(0,T)}^2
\end{equation*}
for every $\delta K\in L^1_{\mathbb R}(0,T)$ satisfying
$K+\delta K\in\mathscr K_T$ and
$\|\delta K\|_{L^1(0,T)}<r_{\mathcal L,\mathfrak K}$
(Corollary~\ref{090926B}).
Error bounds for option prices are not derived in this paper.

In Section \ref{sec:computational} we define a computational framework in which the sensitivity analysis is discretized at multi-factor reference kernels $K_{M,H}$ with $M$ positive
exponential factors approximating the fractional kernel $K_H$. The forward Riccati and backward adjoint equations are solved on a common temporal grid, and the sensitivity kernel and the discrete derivatives are computed by quadrature. In Section \ref{sec:num_validation} the numerical experiments are meant to assess this finite-resolution implementation and to explore the shape of $B_K$, not to establish the analytical results. We first compare the discretized adjoint sensitivities with finite-difference estimates. We then study the lag dependence of the sensitivity kernel and compare the first-order responses to two positive admissible directions with equal $L^1$-mass but different lag profiles, one concentrated at short lags and one diffuse. Next, we evaluate a first-order diagnostic of the effect of the kernel approximation on the Riccati functional. Finally, we examine how these quantities depend on the number of factors and on the spectral range of the multi-factor specification.
 
The paper is organized as follows. Section \ref{sec:analytical_framework} introduces the model, the fixed-$g$ regime, the structural kernel class and the Volterra--Riccati functional. Section \ref{seccion:stability_of_CharcFunction} contains the analytical results: local real-analytic dependence (Section \ref{subsec:existence_stability}), the Fr\'echet derivative and its adjoint representation (Section \ref{subsec:linearized_dual}), and differentiability on the structural class with the quadratic expansion (Section \ref{subsec:quadratic_expansion}). Section \ref{sec:computational} establishes a discretization scheme for the derivative, Section \ref{sec:num_validation} reports the numerical experiments, and Section \ref{sec:conclusions} concludes.

\section{Analytical framework}
\label{sec:analytical_framework}

We first introduce the rough Heston dynamics and specify the fixed-$g$ 
perturbation regime considered throughout the paper. We then define the 
structural kernel class $\mathscr K_T$, the associated Volterra--Riccati 
functional $\mathcal{L}_T$, and the notion of admissible kernel perturbation.
The sensitivity analysis developed in
Section~\ref{seccion:stability_of_CharcFunction}
is formulated entirely at the deterministic level.
The additional conditions ensuring that $\exp(\mathcal{L}_T)$ admits the 
corresponding probabilistic interpretation as the Fourier--Laplace transform of the log-return
are discussed separately in Section~\ref{subsec:probabilistic_interpretation}.

\subsection{Model and fixed input curve}
\label{subsec:model_fixed_g}

Let $T>0$ be a fixed time horizon and let
$(\Omega,\mathcal F,\mathbb F,\mathbb P)$ be a filtered probability
space, where $\mathbb P$ is a pricing measure. We consider the
Volterra dynamics for the discounted asset price $S$ and its stochastic variance $V$ formally given by
\begin{align}
	dS_t &= S_t\sqrt{V_t}\,dW_t, \qquad S_0>0,
	\label{eq:price_dyn}\\[2pt]
	V_t &= g(t) - \lambda\int_0^t K(t-s)V_s\,ds
	+ \nu\int_0^t K(t-s)\sqrt{V_s}\,dB_s,
	\label{eq:variance_dyn}
\end{align}
for $t\in[0,T]$, where $W_t=\rho B_t+\sqrt{1-\rho^{2}}\,B_t^{\perp}$ and
$(B_t,B_t^{\perp})_{t\in [0,T]}$ is a two-dimensional $\mathbb F$-Brownian motion, so
that $d\langle W,B\rangle_t=\rho\,dt$. Conditions under which the system
\eqref{eq:price_dyn}--\eqref{eq:variance_dyn} admits a weak solution are
recalled in Section~\ref{subsec:probabilistic_interpretation}.

The parameter $\lambda>0$ is the damping (mean-reversion) coefficient, $\nu>0$ is the volatility-of-volatility parameter, $\rho\in[-1,1]$ is the instantaneous correlation between the Brownian motions driving the asset price and its variance, and $K$ is the memory kernel that determines the persistence of variance shocks.

The function $g:[0,T]\to\mathbb R_{+}$ is the input variance curve of the model. Following the affine Volterra formulation of
Abi~Jaber and El~Euch~\cite[Eq.~(1.2)]{markovian_rough_volatility}, we
regard $g$ as an exogenous datum and hold it fixed throughout the analysis: when the kernel $K$ is perturbed in Section~\ref{seccion:stability_of_CharcFunction}, the curve $g$ does not vary with $K$. The general framework of affine Volterra processes underlying the Riccati structure is developed by Abi~Jaber, Larsson and Pulido~\cite{abi_jaber_larsson_pulido}.

\begin{remark}
	\label{rem:classical_form}
	The usual rough Heston parametrization describes the variance
	process in terms of its initial value $V_0$, the mean-reversion
	speed $\lambda$, and a mean-reversion level $\theta$, which may
	here depend on time:
	\begin{equation*}
		V_t
		=
		V_0
		+
		\lambda\int_0^t K(t-s)\bigl(\theta(s)-V_s\bigr)\,ds
		+
		\nu\int_0^t K(t-s)\sqrt{V_s}\,dB_s;
	\end{equation*}
	see, for instance, El~Euch and
	Rosenbaum~\cite[Definition~2.1]{Hedging_RH}. These dynamics take
	the form \eqref{eq:variance_dyn} with
	\begin{equation}
		g(t)
		=
		V_0
		+
		\lambda\int_0^t K(t-s)\theta(s)\,ds.
		\label{eq:g_classical}
	\end{equation}
	The formulation in terms of $g$ is therefore a reparametrization of
	the classical model in which the deterministic input curve is taken
	as the primitive datum. We adopt it because the sensitivity analysis
	of Section~\ref{seccion:stability_of_CharcFunction} perturbs $K$
	while keeping $g$ fixed. If $V_0$ and $\theta$ were held fixed
	instead, $g$ would vary with the kernel according
	to~\eqref{eq:g_classical}.
\end{remark}

\subsection{Functional setting and structural kernels}
\label{subsec:structural_kernels}

Throughout the paper, we set
\begin{equation*}
	X:=L^1_{\mathbb R}(0,T)=L^1\bigl((0,T);\mathbb R\bigr),
	\qquad
	Y:=C([0,T];\mathbb C),
\end{equation*}
endowed with the norms
\begin{equation*}
	\|h\|_{L^1(0,T)}
	:=
	\int_0^T|h(s)|\,ds,
	\qquad
	\|f\|_{\infty,[0,T]}
	:=
	\sup_{t\in[0,T]}|f(t)|,
\end{equation*}
respectively. The space $X$ carries the kernel perturbations, and $Y$ carries the Riccati solutions and their linearizations. We write $\mathbb R_+:=[0,\infty)$ and denote by
$L^1_{\mathrm{loc}}(\mathbb R_+)$ the space of measurable functions
that are integrable on $[0,R]$ for every $R>0$.

For Banach spaces $E_1,E_2$ over $\mathbb F\in\{\mathbb R,\mathbb C\}$, we denote by $\mathscr L_{\mathbb F}(E_1,E_2)$ the space of bounded $\mathbb F$-linear operators from $E_1$ to $E_2$; moreover, for a Banach space $E$ and $r>0$, $B_E(0,r):=\{u\in E:\|u\|_E<r\}$ denotes the open ball of radius $r$ centered at the origin.

For $J\in X$ and $f\in Y$, we use the Volterra convolution
\begin{equation}
	(J*f)(t)
	:=
	\int_0^t J(t-s)f(s)\,ds,
	\qquad t\in[0,T],
	\label{eq:volterra_convolution}
\end{equation}
and the same notation for locally integrable $J$ and locally bounded
measurable $f$ on $\mathbb R_+$, with $t\geq0$. If $\mu$ is a locally
finite measure on $\mathbb R_+$ and $f$ is locally bounded and
measurable, we set
\begin{equation*}
	(f*\mu)(t)
	:=
	(\mu*f)(t)
	:=
	\int_{[0,t]}f(t-s)\,\mu(ds),
	\qquad t\geq0.
\end{equation*}

\begin{assumption}
	\label{ass:model}
    The following conditions are considered:
	\begin{enumerate}
		\item[(i)] The kernel $K$ belongs to
		$L^1_{\mathrm{loc}}(\mathbb R_{+})$ and admits a representative,
		still denoted by $K$, which is non-negative, non-increasing and
		continuously differentiable on $(0,\infty)$, with
		\[
		\int_0^t K(s)\,ds>0,
		\qquad t>0.
		\]
		It admits a resolvent of the first kind $L$, given by a
		non-negative locally finite measure on $\mathbb R_{+}$ satisfying
		\[
		(K*L)(t)=(L*K)(t)=1,
		\qquad t>0,
		\]
		and, for every $r\geq0$, the map $s\mapsto L([s,s+r])$ is
		non-increasing on $\mathbb R_{+}$.
		
		\item[(ii)] The parameters satisfy
		\[
		\lambda>0,\qquad \nu>0,\qquad \rho\in[-1,1].
		\]
		
		\item[(iii)] The input curve $g:[0,T]\to\mathbb R_{+}$ is
		continuous.
	\end{enumerate}
\end{assumption}

Assumptions~\ref{ass:model}(ii)--(iii) are maintained throughout,
whereas Assumption~\ref{ass:model}(i) is imposed whenever the kernel
is required to belong to the structural model class. The local
kernel-sensitivity analysis will be carried out under the weaker
requirement that the reference Volterra--Riccati equation admit a
continuous solution, and Assumption~\ref{ass:model}(i) will only be
invoked when identifying the local analytic extension with the
structural Riccati functional.
The kernel conditions in Assumption~\ref{ass:model}(i) are those of
\cite[(2.10)]{abi_jaber_2020}, together with the non-degeneracy
condition $\int_0^t K(s)\,ds>0$, for $t>0$.
Although Assumption~\ref{ass:model}(i) is formulated on $\mathbb R_+$, the deterministic objects considered in this paper up to time $T$ depend on the kernel only through its restriction to $(0,T)$. This motivates the following definition.

\begin{definition}
	\label{def:structural_class}
	We denote by
	\begin{equation*}
		\mathscr K_T
		:=
		\bigl\{
		J\in X:\ J \text{ is the restriction to } (0,T)
		\text{ of a kernel satisfying
			Assumption~\ref{ass:model}(i)}
		\bigr\}
	\end{equation*}
	the structural kernel class on $(0,T)$, regarded as a subset of
	$X$ with the $L^1(0,T)$ topology.
\end{definition}

A kernel in $\mathscr K_T$ may admit several extensions to
$\mathbb R_+$ satisfying Assumption~\ref{ass:model}(i); since the deterministic objects considered in this paper up to time $T$ do not depend on the particular extension, we identify a kernel satisfying Assumption~\ref{ass:model}(i) with its restriction in $\mathscr K_T$ whenever no confusion can arise.

The prototypical example is the fractional kernel
\begin{equation}
	K_H(t)
	=
	\frac{t^{H-1/2}}{\Gamma(H+1/2)},
	\qquad t>0,
	\label{eq:kernel}
\end{equation}
with Hurst parameter $H\in(0,1/2)$, which defines the rough Heston
model of El~Euch and Rosenbaum~\cite{characteristic_RH}. Its
multi-factor approximations are positive finite sums of exponentials,
\begin{equation}
	K_M(t)
	=
	\sum_{i=1}^{M}c_i e^{-\gamma_i t},
	\qquad c_i,\gamma_i>0;
	\label{eq:multifactor_kernel}
\end{equation}
see Abi~Jaber and El~Euch~\cite{abi_jaber_el_euch_2019b}. Both $K_H$
and $K_M$ are completely monotone and satisfy Assumption~\ref{ass:model}(i); see \cite[Example~2.4]{abi_jaber_2020}. Hence, their restrictions to $(0,T)$ belong to $\mathscr K_T$.

\subsection{The Volterra--Riccati functional}
\label{subsec:riccati_functional}

We work on the parameter-independent strip
\[
    \mathcal{Z}
    :=
    \bigl\{z\in\mathbb{C}:\Re z\in[0,1]\bigr\},
\]
without claiming that this domain is maximal. For the parameters fixed
in Assumption~\ref{ass:model}(ii), define the quadratic polynomial
\begin{equation}
	F(z,x)
	=
	\frac12(z^2-z)
	+
	(\rho\nu z-\lambda)x
	+
	\frac{\nu^2}{2}x^2,
	\qquad
	(z,x)\in\mathbb C^2.
	\label{eq:F_def}
\end{equation}
For each kernel $K$ satisfying Assumption~\ref{ass:model}(i) and each
$z\in\mathcal{Z}$, consider the Volterra--Riccati equation
\begin{equation}
	\psi_K(t,z)
	=
	\int_0^t
	K(t-s)\,
	F\bigl(z,\psi_K(s,z)\bigr)\,ds,
	\qquad
	t\in[0,T].
	\label{eq:riccati_a}
\end{equation}
Equation~\eqref{eq:riccati_a} is a time-homogeneous instance of the
Volterra--Riccati equation considered in \cite{abi_jaber_2020}. Under Assumption~\ref{ass:model}(i)--(ii), Theorem~2.5 of \cite{abi_jaber_2020} therefore yields, for every $z\in\mathcal{Z}$, a continuous solution of \eqref{eq:riccati_a} satisfying
\[
    \Re\psi_K(t,z)\leq 0,
    \qquad
    t\in[0,T].
\]
Uniqueness follows from the next lemma, which does not require
Assumption~\ref{ass:model}(i) and is also used in
Section~\ref{seccion:stability_of_CharcFunction} for arbitrary kernels
in $X$.

\begin{lemma}\label{lem:riccati_uniqueness}
	Let $J\in X$ and $z\in\mathbb C$. The
	Volterra--Riccati equation
	\begin{equation*}
		\varphi(t)
		=
		\int_0^t
		J(t-s)F\bigl(z,\varphi(s)\bigr)\,ds,
		\qquad t\in[0,T],
	\end{equation*}
	has at most one solution in $Y$.
\end{lemma}

\begin{proof}
Let $\phi,\widetilde\phi\in Y$ be two solutions and set
$\delta:=\phi-\widetilde\phi$.
Since $F(z,\cdot)$ is locally Lipschitz and both solutions are bounded,
there exists $M\ge0$ such that
\[
|\delta(t)|\le M\int_0^t|J(t-s)|\,|\delta(s)|\,ds,
\qquad t\in[0,T].
\]
Choose $\tau\in(0,T]$ so that
$q:=M\int_0^\tau|J(u)|\,du<1$.
On $[0,\tau]$, this gives
$\|\delta\|_{\infty,[0,\tau]}
\le q\|\delta\|_{\infty,[0,\tau]}$, hence $\delta=0$.
If $\delta=0$ on $[0,a]$, the same estimate on
$[a,\min\{a+\tau,T\}]$ gives $\delta=0$ there as well.
A finite iteration yields $\delta=0$ on $[0,T]$.
\end{proof}

By Lemma~\ref{lem:riccati_uniqueness}, the solution of
\eqref{eq:riccati_a} provided by \cite[Theorem~2.5]{abi_jaber_2020} is
the unique solution in $Y$. Thus, $\psi_K(\cdot,z)\in Y$ is well defined for every $z\in\mathcal Z$. Since \eqref{eq:riccati_a} involves $K$ only on $(0,T)$, the solution $\psi_K(\cdot,z)$ depends only on the restriction of $K$ to $(0,T)$, and $\psi_J(\cdot,z)$ is well defined for every $J\in\mathscr K_T$.

We define the deterministic functional
\begin{equation}
	\mathcal{L}_T(J,z)
	=
	\int_0^T
	F\bigl(z,\psi_J(T-r,z)\bigr)\,g(r)\,dr,
	\qquad
	(J,z)\in\mathscr{K}_T\times\mathcal{Z},
	\label{eq:logchar_conv}
\end{equation}
which, by the same observation, is a well-defined map
$\mathcal L_T(\cdot,z):\mathscr K_T\to\mathbb C$ for every
$z\in\mathcal Z$. By the change of variables $s=T-r$,
\begin{equation}
	\mathcal{L}_{T}(J,z)
	=
	\int_0^T
	F\bigl(z,\psi_J(s,z)\bigr)\,g(T-s)\,ds.
	\label{eq:functional_LT_alt}
\end{equation}
This representation will be used throughout Section~\ref{seccion:stability_of_CharcFunction}, where the local dependence of the functional on the kernel is studied with $\lambda,\nu,\rho$ and $g$ held fixed.

\subsection{Admissible kernel perturbations}
\label{subsec:admissible_perturbations}

Kernel perturbations are measured in the norm of $X$.
The structural class $\mathscr{K}_T$ has empty interior in $X$:
for any $K\in\mathscr{K}_T$ and $0<\delta<T$, the function
$J_\delta:=K-(K+1)\mathbf{1}_{(0,\delta)}$ is negative on
$(0,\delta)$, whereas
$\|J_\delta-K\|_{L^1(0,T)}
=\int_0^\delta (K(s)+1)\,ds\to0$ as $\delta\downarrow0$.
Thus arbitrarily small $L^1$-perturbations of a structural
kernel need not remain in $\mathscr{K}_T$.
This motivates the following notion of admissible
kernel perturbation.

\begin{definition}\label{def:admissible_direction}
    Let $K\in\mathscr{K}_T$. A direction $h\in X$ is called admissible at $K$ if there exists $\varepsilon_0 > 0$ such that
    \begin{equation*}
        K+\varepsilon h\in\mathscr{K}_T
        \qquad
        \text{for every }\varepsilon\in [0,\varepsilon_0].
    \end{equation*}
    The elements $\delta K = \varepsilon h$, $\varepsilon\in [0,\varepsilon_0]$, are called the corresponding admissible increments at $K$.
\end{definition}

Admissibility is a condition on the deterministic class $\mathscr{K}_T$ only; it does not involve the probabilistic assumptions of Section~\ref{subsec:probabilistic_interpretation}.

\begin{remark}
	\label{rem:kernel_examples}
	Let $K_M$ be the multi-factor kernel~\eqref{eq:multifactor_kernel}.
	Every direction $h\in X$ obtained by restricting to $(0,T)$ a function of the form
	\[
    	h(t)=\sum_{i=1}^{M}d_i e^{-\gamma_i t},
        \qquad t>0,
    	\qquad d_i\in\mathbb{R},
	\]
	is admissible at $K_M$. Indeed, since $M<\infty$ and $c_i > 0$, one can choose $\varepsilon_0 >0$ such that $c_i + \varepsilon d_i > 0$ for every $i=1,\ldots,M$ and $\varepsilon\in [0,\varepsilon_0]$. Hence,
	\[
	    K_M(t)+\varepsilon h(t)
    	=
    	\sum_{i=1}^{M}(c_i+\varepsilon d_i)e^{-\gamma_i t}
	\]
	is a positive completely monotone kernel and satisfies Assumption~\ref{ass:model}(i); see \cite[Example~2.4]{abi_jaber_2020}.
\end{remark}

\subsection{Probabilistic interpretation}
\label{subsec:probabilistic_interpretation}

The following additional assumptions are imposed only to give $\mathcal L_T(K,z)$ its probabilistic interpretation in the Volterra--Heston model.

For $\vartheta\geq0$ and a function $f$ on $\mathbb R_+$, we write
$\Delta_\vartheta f:=f(\cdot+\vartheta)$. If $f$ is right-continuous
and of locally bounded variation, $df$ denotes the associated Lebesgue--Stieltjes measure.

\begin{assumption}
    \label{ass:stochastic}
    The kernel $K$ satisfies Assumption~\ref{ass:model}(i), with resolvent of the first kind $L$, and there exists $\kappa\in(0,2]$ such that:
    \begin{enumerate}
        \item[(i)] $K\in L^2_{\mathrm{loc}}(\mathbb R_+, \mathbb{R})$ and,
        \[
            \int_0^\vartheta K(s)^2\,ds
            =
            O(\vartheta^\kappa)
            \quad \text{and}\quad
            \int_0^T
            \bigl(K(s+\vartheta)-K(s)\bigr)^2\,ds
            =
            O(\vartheta^\kappa)
            \quad\text{for every } T<\infty \text{ as } \vartheta\downarrow0.
        \]

        \item[(ii)] The input curve $g$ is the restriction to $[0,T]$ of a function $g:\mathbb R_+\to\mathbb R$, still denoted by $g$, which is locally H\"older continuous of every order strictly smaller than $\kappa/2$, satisfies $g(0)\geq0$, and
        \begin{equation}
            \Delta_\vartheta g
            -
            (\Delta_\vartheta K*L)(0)\,g
            -
            d(\Delta_\vartheta K*L)*g
            \geq0,
            \qquad
            \vartheta\geq0.
            \label{180826}
        \end{equation}

        \item[(iii)] For every $\vartheta\in(0,1]$, the translated kernel
        $\Delta_\vartheta K$ is not identically zero and admits a non-negative resolvent of the first kind $L_\vartheta$ such that $s\mapsto L_\vartheta([s,s+r])$ is non-increasing for every $r\geq 0$.
    \end{enumerate}
\end{assumption}

Under Assumption~\ref{ass:model}(i), the function $\Delta_{\vartheta}K\ast L$ is right-continuous and non-decreasing \cite[Remark~B.3]{markovian_rough_volatility}, so the measure $d(\Delta_{\vartheta} K\ast L)$ in \eqref{180826} is well defined. Moreover, Assumption~\ref{ass:stochastic}(ii) holds, in particular, for every non-decreasing $g$ with $g(0)\geq 0$ and the required H\"older regularity; see \cite[Example~2.2]{markovian_rough_volatility}.

\begin{proposition}
    \label{prop:probabilistic_interpretation}
    Under Assumption~\ref{ass:stochastic}, the system
    \eqref{eq:price_dyn}--\eqref{eq:variance_dyn} admits a continuous
    weak solution $(S,V)$ with values in $\mathbb R_+^2$, and
    \begin{equation}
        \Phi_K(z)
        :=
        \mathbb E\!\left[
            \exp\!\left(z\log(S_T/S_0)\right)
        \right]
        =
        \exp\bigl(\mathcal L_T(K,z)\bigr),
        \qquad z\in\mathcal Z.
        \label{eq:char_fn}
    \end{equation}
    In particular,
    $\mathbb E[S_T/S_0]=1$.
\end{proposition}

\begin{proof}
    Assumption~\ref{ass:stochastic}(i) is condition $(H_0)$ of \cite{markovian_rough_volatility}, while
    Assumption~\ref{ass:model}(i) gives condition $(H_1)$ of the same reference for the kernel $K$. Moreover, Assumption~\ref{ass:stochastic}(ii) is precisely the admissibility condition on the input curve appearing in
    \cite[(2.4)--(2.5)]{markovian_rough_volatility}. Hence, weak existence follows from \cite[Theorem~2.1]{markovian_rough_volatility}.

    For the transform formula, Assumption~\ref{ass:stochastic}(iii),
    together with Assumption~\ref{ass:model}(i), gives condition $(H_1)$
    for the translated kernels required in
    \cite[Theorem~2.3]{markovian_rough_volatility}.
    
    Applying \cite[Theorem~2.5]{abi_jaber_2020} on each finite horizon $[0,n]$, $n\in\mathbb N$, and using Lemma~\ref{lem:riccati_uniqueness} to identify the solutions on overlapping intervals, the continuous solution $\psi_K(\cdot,z)$ extends uniquely to a continuous solution on
    $\mathbb R_+$. In particular, $\psi_K(\cdot,z)\in L^2_{\mathrm{loc}}(\mathbb R_+)$.
    
    Specializing \cite[Theorem~2.3]{markovian_rough_volatility} to the Fourier--Laplace transform of the log-price alone, with $u=(z,0)$
    and $f\equiv0$, one has $\psi_1\equiv z$, while the second component
    $\psi_2\in L^2_{\mathrm{loc}}(\mathbb R_+)$ is the unique global
    solution of \eqref{eq:riccati_a}. Hence
    \[
        \psi_2=\psi_K(\cdot,z).
    \]
    Moreover, the admissibility conditions of
    \cite[Theorem~2.3]{markovian_rough_volatility} reduce here to
    $\Re z\in[0,1]$.
    
    Therefore,
    \[
        \mathbb E\!\left[
            \exp\!\left(z\log(S_T/S_0)\right)
        \right]
        =
        \exp\bigl(\mathcal L_T(K,z)\bigr),
        \qquad z\in\mathcal Z.
    \]

    Finally, for $z=1$, $F(1,0)=0$, so uniqueness gives
    $\psi_K(\cdot,1)\equiv0$. Hence
    $\mathcal L_T(K,1)=0$, and therefore
    $\mathbb E[S_T/S_0]=1$.
\end{proof}

For $v\in\mathbb R$, $\Phi_K(iv)$ is the characteristic function of
$\log(S_T/S_0)$, so $\mathcal L_T(K,iv)$ is the corresponding
characteristic exponent. For admissible perturbations $K+\delta K$ not known to satisfy Assumption~\ref{ass:stochastic}, no probabilistic interpretation is asserted; the sensitivity analysis in Section~\ref{seccion:stability_of_CharcFunction} treats $\mathcal L_T$ solely as a deterministic functional.

\section{Kernel sensitivity and adjoint representation}
\label{seccion:stability_of_CharcFunction}

We study the local dependence of the deterministic Riccati functional
$\mathcal L_T(K,z)$ defined in \eqref{eq:functional_LT_alt} on the
memory kernel. Throughout this section, the parameters in
Assumption~\ref{ass:model}(ii) and the input curve $g$ in
Assumption~\ref{ass:model}(iii) are held fixed. Notice that holding $g$ fixed is not equivalent to holding the forward variance curve $t\mapsto\mathbb{E}[V_t]$ fixed.

The perturbation variable $\eta$ used in this section ranges over the ambient
space $X$ and is not required to be admissible in
the sense of Definition~\ref{def:admissible_direction}. This enlargement 
provides the open domain needed for the implicit function theorem. 
Admissibility enters only when the resulting analytic extension is 
identified with the structural Riccati functional: every admissible 
increment $\delta K$ is an increment $\eta$ for which $K+\eta$ admits an 
extension satisfying Assumption~\ref{ass:model}(i).

\begin{assumption}
	\label{ass:reference_riccati}
	The reference kernel satisfies
	\begin{equation*}
		K^{\mathrm{ref}}\in X,
	\end{equation*}
	and, for every $z\in\mathcal Z$, the Volterra--Riccati equation
	\begin{equation}
		\psi_{K^{\mathrm{ref}}}(t,z)
		=
		\int_0^t
		K^{\mathrm{ref}}(t-s)F\bigl(z,\psi_{K^{\mathrm{ref}}}
        (s,z)\bigr)\,ds,
		\qquad t\in[0,T],
		\label{eq:reference_riccati}
	\end{equation}
	admits a solution
	$\psi_{K^{\mathrm{ref}}}(\cdot,z)\in Y$.
\end{assumption}

Under the parameter restrictions of
Assumption~\ref{ass:model}(ii), every kernel satisfying
Assumption~\ref{ass:model}(i) also satisfies
Assumption~\ref{ass:reference_riccati}, by
\cite[Theorem~2.5]{abi_jaber_2020}. The additional structural
conditions imposed in Assumption~\ref{ass:model}(i), however, are not
needed once a continuous reference solution is available. Uniqueness is 
not included in Assumption~\ref{ass:reference_riccati}; it is proved
in Lemma~\ref{lem:riccati_uniqueness} for arbitrary
kernels in $X$ by a direct local contraction
argument.

The section is organized as follows. We first establish local well 
posedness and construct an analytic extension of the Riccati solution with 
respect to kernel perturbations. We then derive its first-order 
sensitivity through a dual Volterra representation, and finally obtain a 
quadratic expansion of the corresponding Riccati functional, uniformly for 
$z$ in compact subsets of $\mathcal Z$.

\subsection{Local real-analytic extension}
\label{subsec:existence_stability}

When applying the real-analytic implicit function theorem, we regard $Y$ as a real Banach space by restricting scalar multiplication from $\mathbb C$ to $\mathbb R$; the underlying set, vector addition and norm remain unchanged. Likewise, the parameter space $\mathbb C$ is regarded as the two-dimensional real Banach space $\mathbb R^2$. This real formulation is natural because the kernel perturbations belong to the real space $X$. The linearized operators introduced in this subsection are nevertheless $\mathbb C$-linear on $Y$.

We next consider the linearization of equation  \eqref{eq:reference_riccati}. For each $z\in \mathcal{Z}$, we define
\begin{equation} \label{131026}
q_{K^{\mathrm{ref}},z}(s):=\partial _{x}F(z,\psi_{K^{\mathrm{ref}}}(s,z))=
(\rho \nu z-\lambda )+\nu
^{2}\psi _{K^{\mathrm{ref}}}(s,z),\qquad s\in \lbrack 0,T],
\end{equation}%
and the linearized Volterra operator
\begin{equation}
(\mathcal{A}_{K^{\mathrm{ref}},z}\varphi )(t):=\int_{0}^{t}K^{\mathrm{ref}}
(t-s)q_{K^{\mathrm{ref}},z}(s)\varphi
(s)ds,\qquad t\in \lbrack 0,T].  
\label{eq:linearized_operator}
\end{equation}
The following result shows that $\mathcal A_{K^{\mathrm{ref}},z}$ is 
bounded and that $I-\mathcal A_{K^{\mathrm{ref}},z}$ is boundedly 
invertible.

\begin{proposition}
	\label{prop:linearized_invertibility}
	Under Assumption~\ref{ass:reference_riccati}, for every 
    $z\in\mathcal{Z}$, the operator $\mathcal A_{K^{\mathrm{ref}},z}$ is a 
    bounded $\mathbb C$-linear operator on $Y$, and $I-\mathcal 
    A_{K^{\mathrm{ref}},z}$ is boundedly invertible on $Y$.
\end{proposition}

\begin{proof}
	Now fix $z\in\mathcal{Z}$. By Assumption~\ref{ass:reference_riccati},
	$q_{K^{\mathrm{ref}},z}\in C([0,T];\mathbb C)$. Let 
    $\overline{K^{\mathrm{ref}}}$ denote the extension of 
    $K^{\mathrm{ref}}$ by zero outside $(0,T)$ and define
	\begin{equation*}
		a_{K^{\mathrm{ref}},z}(t,s)
		:=
		\overline{K^{\mathrm{ref}}}(t-s)q_{K^{\mathrm{ref}},z}(s),
		\qquad
		(t,s)\in[0,T]^2.
	\end{equation*}
	Since $\overline{K^{\mathrm{ref}}}(t-s)=0$ for $s\geq t$, equation
	\eqref{eq:linearized_operator} can equivalently be written as
	\begin{equation*}
		(\mathcal A_{K^{\mathrm{ref}},z}\varphi)(t)
		=
		\int_0^T a_{K^{\mathrm{ref}},z}(t,s)\varphi(s)\,ds.
	\end{equation*}

	For $t,t_0\in[0,T]$,
	\begin{align}
		\|a_{K^{\mathrm{ref}},z}(t,\cdot)-a_{K^{\mathrm{ref}},z}
        (t_0,\cdot)\|_{L^1(0,T)}
		&\leq
		\|q_{K^{\mathrm{ref}},z}\|_{\infty,[0,T]}
		\bigl\|
			\overline{K^{\mathrm{ref}}}(t-\cdot)
			-
			\overline{K^{\mathrm{ref}}}(t_0-\cdot)
		\bigr\|_{L^1(\mathbb R)}.
		\label{eq:continuous_type_estimate}
	\end{align}
	Translations are strongly continuous on $L^1(\mathbb R)$, that is, for 
    every $f\in L^1(\mathbb R)$,
	\begin{equation*}
		\|f(\,\cdot+h)-f\|_{L^1(\mathbb R)}
		\longrightarrow0
		\qquad\text{as }h\to0
	\end{equation*}
	(see \cite[Lemma 4.3]{brezis2011functional}). Consequently, the right-hand side of \eqref{eq:continuous_type_estimate} tends to zero as
	$t\rightarrow t_0$. Since $[0,T]$ is compact, this $L^1$-continuity is 
    uniform.

	It follows in particular that, for every $\varphi\in Y$,
	\begin{equation*}
		|(\mathcal{A}_{K^{\mathrm{ref}},z}\varphi)(t)
		-
		(\mathcal{A}_{K^{\mathrm{ref}},z}\varphi)(t_0)|
		\leq
		\|\varphi\|_{\infty,[0,T]}
		\|a_{K^{\mathrm{ref}},z}(t,\cdot)-a_{K^{\mathrm{ref}},z}
        (t_0,\cdot)\|_{L^1(0,T)}
		\longrightarrow0.
	\end{equation*}
	Thus, $\mathcal{A}_{K^{\mathrm{ref}},z}$ maps $Y$ into itself. 
    Moreover,
	\begin{equation*}
		\|\mathcal{A}_{K^{\mathrm{ref}},z}\varphi\|_{\infty,[0,T]}
		\leq
		\|K^{\mathrm{ref}}\|_{L^1(0,T)}
		\|q_{K^{\mathrm{ref}},z}\|_{\infty,[0,T]}
		\|\varphi\|_{\infty,[0,T]},
	\end{equation*}
	so $\mathcal{A}_{K^{\mathrm{ref}},z}$ is a bounded $\mathbb{C}$-linear
    operator on $Y$.

	The estimate \eqref{eq:continuous_type_estimate}, together with
	\begin{equation*}
		\sup_{t\in[0,T]}
		\|a_{K^{\mathrm{ref}},z}(t,\cdot)\|_{L^1(0,T)}
		\leq
		\|K^{\mathrm{ref}}\|_{L^1(0,T)}
		\|q_{K^{\mathrm{ref}},z}\|_{\infty,[0,T]},
	\end{equation*}
	shows that $a_{K^{\mathrm{ref}},z}$ is a Volterra kernel of continuous 
    type in the sense of \cite[Definition~9.5.2]{gripenberg}. Together 
    with the preceding uniform bound, the uniform $L^1$-continuity shows 
    that $a_{K^{\mathrm{ref}},z}$, and hence also $-
    a_{K^{\mathrm{ref}},z}$, is a Volterra kernel of bounded uniformly 
    continuous type. By \cite[Theorem~9.5.5(ii)]{gripenberg}, $-
    a_{K^{\mathrm{ref}},z}$ has a resolvent of bounded uniformly 
    continuous type on $[0,T]$, which is in particular of type $L^\infty$. 
    Since every $f\in Y=C([0,T];\mathbb C)$ is bounded and uniformly 
    continuous, \cite[Theorem~9.5.4]{gripenberg} yields a unique 
    $\varphi\in Y$ satisfying
    \begin{equation*}
        (I-\mathcal A_{K^{\mathrm{ref}},z})\varphi=f.
    \end{equation*}
    Hence $I-\mathcal{A}_{K^{\mathrm{ref}},z}$ is bijective on $Y$, and 
    the bounded inverse theorem shows that its inverse is bounded. Thus, 
    $I-\mathcal{A}_{K^{\mathrm{ref}},z}$ is boundedly invertible.
\end{proof}

All analyticity and differentiation statements are understood
with respect to these real Banach-space structures. More
precisely, a map between real Banach spaces is called real-analytic if,
in a neighborhood of each point, it admits a norm-convergent expansion
in continuous homogeneous polynomials; see
\cite[Sections~4.1--4.3]{buffoni_toland}.

Let $E$ be a real Banach space, and let
$\Phi:\mathcal{U}\rightarrow E$ be defined on an open subset 
$\mathcal{U}\subset X\times\mathbb{C}$.
Whenever $\Phi$ is partially Fr\'echet differentiable in its
first variable at $(J,z)\in\mathcal U$, we denote by
\begin{equation*}
	D_\eta\Phi(J,z)
	\in
	\mathscr L_{\mathbb R}(X,E)
\end{equation*}
this derivative. 
Thus, $D_\eta\Phi(J,z)$ is the unique bounded $\mathbb{R}$-linear
operator such that
\begin{equation*}
	\frac{
		\|\Phi(J+h,z)-\Phi(J,z)-D_\eta\Phi(J,z)[h]\|_E
	}{
		\|h\|_{L^1(0,T)}
	}
	\longrightarrow 0
\end{equation*}
as $h\rightarrow0$ in $X$. We use $D_\eta^2\Phi(J,z)$ for the 
corresponding second partial  Fr\'{e}chet derivative, viewed as a bounded
$\mathbb{R}$-bilinear map from $X\times X$ to $E$. These notions and
the associated chain rule are recalled in \cite[Sections~3.1--3.2]
{buffoni_toland}.

In what follows, we extend the map $\psi_{K^{\mathrm{ref}}}(\cdot ,z)$ 
real-analytically with respect to a kernel perturbation and the parameter 
$z$.

\begin{theorem}
	\label{thm:local_analytic_solution}
	Under Assumption~\ref{ass:reference_riccati}, the following statements 
    hold.

    \begin{itemize}
        \item[(i)] There exists an open set
        \begin{equation*}
            \{0\}\times \mathcal{Z}\subset 
            \mathcal{O}_{K^{\mathrm{ref}}\times \mathcal{Z}}\subset X\times
            \mathbb{C}
        \end{equation*}%
        and a real-analytic map
        \begin{equation*}
            \Psi _{K^{\mathrm{ref}}}:\mathcal{O}_{K^{\mathrm{ref}}\times 
            \mathcal{Z}}\longrightarrow Y
        \end{equation*}%
        such that
        \begin{equation}
            \Psi_{K^{\mathrm{ref}}}(\eta,z)=(K^{\mathrm{ref}}+\eta)\ast 
            F(z,\Psi_{K^{\mathrm{ref}}}(\eta,z)),\qquad (\eta,z)\in 
            \mathcal{O}%
            _{K^{\mathrm{ref}}\times \mathcal{Z}},  
            \label{eq:extended_riccati}
        \end{equation}%
        and
        \begin{equation}
            \Psi_{K^{\mathrm{ref}}}(0,z)=\psi_{K^{\mathrm{ref}}}
            (\cdot,z),\qquad z\in \mathcal{Z}.
            \label{eq:extended_solution_at_reference}
        \end{equation}

        \item[(ii)] Whenever $z\in \mathcal{Z}$, $(\eta,z)\in 
        \mathcal{O}_{K^{\mathrm{ref}}\times \mathcal{Z}}
        $, and $K^{\mathrm{ref}}+\eta$ is the restriction to $(0,T)$ of a 
        kernel satisfying Assumption~%
        \ref{ass:model}(i), one has
        \begin{equation}
            \Psi _{K^{\mathrm{ref}}}(\eta,z)=\psi _{K^{\mathrm{ref}}+\eta}
            (\cdot ,z).
            \label{eq:extended_solution_coincidence}
        \end{equation}

        \item[(iii)] In particular, the set
        \begin{equation*}
            \mathcal{U}_{K^{\mathrm{ref}}}:=\{z\in \mathbb{C}:(0,z)\in 
            \mathcal{O}_{K^{\mathrm{ref}}\times \mathcal{Z}%
            }\}
        \end{equation*}%
        is an open neighborhood of $\mathcal{Z}$ in $\mathbb{C}$, and
        \begin{equation*}
            \widehat{\psi }_{K^{\mathrm{ref}}}(z):=\Psi _{K^{\mathrm{ref}}}
            (0,z),\qquad z\in \mathcal{U}_{K^{\mathrm{ref}}},
        \end{equation*}%
        defines a real-analytic extension of $z\mapsto \psi 
        _{K^{\mathrm{ref}}}(\cdot ,z)$ from $%
        \mathcal{Z}$ to $\mathcal{U}_{K^{\mathrm{ref}}}$.
    \end{itemize}
\end{theorem}

\begin{proof}
	(i) For $z\in\mathbb C$ and $\varphi\in Y$, we use the Nemytskii
	notation
	\begin{equation*}
		F(z,\varphi)(t)
		:=
		F\bigl(z,\varphi(t)\bigr),
		\qquad t\in[0,T].
	\end{equation*}
	Thus,
	\begin{equation*}
		F(z,\varphi)
		=
		\frac12(z^2-z)\mathbf 1
		+
		(\rho\nu z-\lambda)\varphi
		+
		\frac{\nu^2}{2}\varphi^2,
	\end{equation*}
	where $\mathbf{1}$ denotes the constant function equal to one and
	$\varphi^{2}$ is understood pointwise.

	Define
	\begin{equation*}
		\mathscr{G}:
		X\times\mathbb{C}\times Y
		\longrightarrow Y
	\end{equation*}
	by
	\begin{equation}
		\mathscr{G}(\eta,z,\varphi)
		:=
		\varphi
		-
		(K^{\mathrm{ref}}+\eta)*F(z,\varphi).
		\label{eq:implicit_mapping}
	\end{equation}
	This map is well defined. Indeed, the Volterra convolution
	\begin{equation*}
		C:X\times Y\longrightarrow Y,
		\qquad
		C(J,f):=J*f,
	\end{equation*}
	is continuous and bilinear: the translation argument used in the
	proof of Proposition~\ref{prop:linearized_invertibility} shows that
	$J\ast f$ is continuous, while
	\begin{equation}
		\|J\ast f\|_{\infty,[0,T]}
		\leq
		\|J\|_{L^1(0,T)}
		\|f\|_{\infty,[0,T]}.
		\label{eq:volterra_convolution_bound}
	\end{equation}
	The pointwise product
	\begin{equation*}
		Y\times Y\longrightarrow Y,
		\qquad
		(\varphi_{1},\varphi_{2})\longmapsto\varphi_{1}\varphi_{2},
	\end{equation*}
	is also continuous and bilinear, since
	\begin{equation*}
		\|\varphi_1\varphi_2\|_{\infty,[0,T]}
		\leq
		\|\varphi_1\|_{\infty,[0,T]}
		\|\varphi_2\|_{\infty,[0,T]}.
	\end{equation*}
	Consequently, $\mathscr G$ is a finite sum of continuous
	multilinear expressions in the real variables
	$(\eta,z,\varphi)$ and is therefore real-analytic; see
	\cite[Sections~4.1--4.3]{buffoni_toland}.

	For every $z_0\in\mathcal{Z}$, Assumption
	\ref{ass:reference_riccati} gives
	\begin{equation}
		\mathscr G
		\bigl(0,z_{0},\psi_{K^{\mathrm{ref}}}(\cdot,z_{0})\bigr)
		=0.
		\label{eq:implicit_base_point}
	\end{equation}
	For $w\in Y$, differentiation of
	\eqref{eq:implicit_mapping} with respect to $\varphi $ gives
	\begin{equation*}
		D_\varphi\mathscr G(\eta,z,\varphi)[w]
		=
		w
		-
		(K^{\mathrm{ref}}+\eta)\ast
		\left(
			\partial_xF(z,\varphi)w
		\right).
	\end{equation*}
	In particular, by the definition
	\eqref{eq:linearized_operator} of $\mathcal A_{K^{\mathrm{ref}},z_0}$,
	\begin{equation}
		D_\varphi\mathscr G
		\bigl(0,z_{0},\psi_{K^{\mathrm{ref}}}(\cdot,z_{0})\bigr)
		=
		I-\mathcal A_{K^{\mathrm{ref}},z_{0}}.
		\label{eq:implicit_state_derivative_at_base}
	\end{equation}
	This operator is boundedly invertible by
	Proposition~\ref{prop:linearized_invertibility}.

	The real-analytic implicit function theorem in Banach spaces
	\cite[Theorem~4.5.4]{buffoni_toland} therefore provides, for every
	$z_{0}\in\mathcal{Z}$, an open neighborhood
	\begin{equation*}
		(0,z_{0})\in \mathcal{U}_{K^{\mathrm{ref}}\times z_{0}}
        \subset X\times \mathbb{C}
	\end{equation*}
	and a real-analytic map
	\begin{equation*}
		\Psi_{K^{\mathrm{ref}},z_{0}}:\mathcal{U}_{K^{\mathrm{ref}}\times 
        z_{0}}\longrightarrow Y
	\end{equation*}
	such that
	\begin{equation*}
		\mathscr G
		\bigl(\eta,z,\Psi_{K^{\mathrm{ref}},z_{0}}(\eta,z)\bigr)=0
	\end{equation*}
	for every $(\eta,z)\in\mathcal{U}_{K^{\mathrm{ref}}\times z_0}$ and
	\begin{equation*}
		\Psi_{K^{\mathrm{ref}},z_{0}}(0,z_{0})=\psi_{K^{\mathrm{ref}}}
        (\cdot,z_{0}).
	\end{equation*}
    
    On the other hand, given $z_{0}, z_{1}\in\mathcal{Z}$, $z_{0}\neq 
    z_{1}$ such that $\mathcal{U}_{K^{\mathrm{ref}}\times z_{0}}\cap 
    \mathcal{U}_{K^{\mathrm{ref}}\times z_{1}}\neq \emptyset $, for every $(\eta,z)\in \mathcal{U}_{K^{\mathrm{ref}}\times z_{0}}\cap\mathcal{U}_{K^{\mathrm{ref}}\times z_{1}}$, both $\Psi_{K^{\mathrm{ref}},z_{0}}(\eta,z)$ and $\Psi _{K^{\mathrm{ref}},z_{1}}(\eta,z)$ solve
    \begin{equation*}
        \varphi =(K^{\mathrm{ref}}+\eta)\ast F(z,\varphi ).
    \end{equation*}%
	Since $K^{\mathrm{ref}}+\eta\in L^1_{\mathbb R}(0,T)$, the uniqueness 
    assertion of Lemma~\ref{lem:riccati_uniqueness}, applied with 
    $J=K^{\mathrm{ref}}+\eta$, shows that both maps necessarily coincide,
	\begin{equation*}
        \Psi_{K^{\mathrm{ref}},z_{0}}(\eta,z)
        =
        \Psi_{K^{\mathrm{ref}},z_{1}}(\eta,z)\text{ on 
        }\mathcal{U}_{K^{\mathrm{ref}}\times
        z_{0}}\cap \mathcal{U}_{K^{\mathrm{ref}}\times z_{1}}.
    \end{equation*}
    
	Hence, the local maps agree on all overlaps. Setting
	\begin{equation*}
        \mathcal{O}_{K^{\mathrm{ref}}\times \mathcal{Z}}
        :=
        \bigcup_{z\in \mathcal{Z}}\mathcal{U}%
        _{K^{\mathrm{ref}}\times z},
    \end{equation*}%
	we can define a unique map
	\begin{equation*}
        \Psi_{K^{\mathrm{ref}}}:\mathcal{O}_{K^{\mathrm{ref}}\times 
        \mathcal{Z}}\longrightarrow Y.
    \end{equation*}%
	The set $\mathcal{O}_{K^{\mathrm{ref}}\times \mathcal{Z}}$ is open and 
    contains $\{0\}\times \mathcal{Z}$, while $\Psi _{K^{\mathrm{ref}}}$ 
    is real-analytic because it agrees locally with real-analytic maps 
    $\Psi _{K^{\mathrm{ref}},z}$.

	(ii) For $z\in\mathcal{Z}$, both $\Psi_{K^{\mathrm{ref}}}(0,z)$ and
	$\psi_{K^{\mathrm{ref}}}(\cdot,z)$ solve the reference equation
	\eqref{eq:reference_riccati}. Lemma
	\ref{lem:riccati_uniqueness} therefore gives
	\eqref{eq:extended_solution_at_reference}. The same argument,
	applied to $J=K^{\mathrm{ref}}+\eta$, proves
	\eqref{eq:extended_solution_coincidence} whenever its stated
	hypotheses hold.

    (iii) The set $\mathcal{U}_{K^{\mathrm{ref}}}$ defined in the 
    statement is open because it is the inverse image of 
    $\mathcal{O}_{K^{\mathrm{ref}}\times\mathcal{Z}}$ under the continuous embedding $z\longmapsto (0,z)$.
    The restriction $z\mapsto \Psi_{K^{\mathrm{ref}}}(0,z)$ is real-analytic on $\mathcal{U}_{K^{\mathrm{ref}}}$ and, by 
    \eqref{eq:extended_solution_at_reference}, agrees with $\psi 
    _{K^{\mathrm{ref}}}(\cdot ,z)$ for $z\in \mathcal{Z}$. This proves the
    asserted extension in the transform parameter.
\end{proof}

As a consequence of the analyticity, the map $\Psi _{K^{\mathrm{ref}}}$ is
Fr\'{e}chet differentiable on $\mathcal{O}_{K^{\mathrm{ref}}\times 
\mathcal{Z}}$, and we can characterize explicitly its partial derivative 
with respect to the kernel variable.

\begin{proposition} \label{prop:differentiability}
Suppose that Assumption~\ref{ass:reference_riccati} holds. For every $z\in 
\mathcal{Z}$, $\Psi _{K^{\mathrm{ref}}}$ is differentiable on $\mathcal{%
O}_{K^{\mathrm{ref}}\times \mathcal{Z}}$, and its partial Fr\'{e}chet 
derivative
\begin{equation*}
D_{\eta}\Psi _{K^{\mathrm{ref}}}(0,z)\in \mathscr L_{\mathbb{R}}(X,Y)
\end{equation*}%
is given by
\begin{equation}
D_{\eta}\Psi_{K^{\mathrm{ref}}}(0,z)[h]
=
(I-\mathcal{A}_{K^{\mathrm{ref}},z})^{-1}\left( h\ast F(z,\psi 
_{K^{\mathrm{ref}}}(\cdot ,z))\right) ,\qquad h\in X.  
\label{eq:derivative_solution_map}
\end{equation}
\end{proposition}

\begin{proof}
By Theorem~\ref{thm:local_analytic_solution},
$\Psi_{K^{\mathrm{ref}}}$ is real-analytic.
Differentiating \eqref{eq:extended_riccati} with respect to $\eta$
at $\eta=0$ in the direction $h\in X$, and using
\eqref{eq:extended_solution_at_reference}, gives
\[
(I-\mathcal A_{K^{\mathrm{ref}},z})
D_\eta\Psi_{K^{\mathrm{ref}}}(0,z)[h]
=h*F\bigl(z,\psi_{K^{\mathrm{ref}}}(\cdot,z)\bigr).
\]
Invertibility from Proposition~\ref{prop:linearized_invertibility}
yields \eqref{eq:derivative_solution_map}.
\end{proof}

We establish two further properties of $\Psi_{K^{\mathrm{ref}}}$. In 
particular, its first-order Taylor expansion at $\eta=0$ has a quadratic
remainder, uniformly for $z$ in compact subsets of $\mathcal Z$; moreover, 
$\Psi_{K^{\mathrm{ref}}}$ vanishes identically at the endpoints $z\in\{0,1\}$.

\begin{proposition} \label{prop:compacity}
Under Assumption~\ref{ass:reference_riccati}, the following
statements hold:

\begin{itemize}
\item[(i)] For every compact set $\mathfrak{K}\subset \mathcal{Z}$, there
exist constants $r_{\psi, \mathfrak{K}}, \ C_{\psi ,\mathfrak{K}}>0$ such that
\begin{equation}
    B_{X}(0,r_{\psi, \mathfrak{K}})\times \mathfrak{K}\subset \mathcal{O}_{K^{\mathrm{ref}}\times
    \mathcal{Z}},  
    \label{eq:uniform_solution_neighborhood}
\end{equation}%
where $\mathcal{O}_{K^{\mathrm{ref}}\times \mathcal{Z}}$ is the open set defined in Theorem \ref{thm:local_analytic_solution} and
\begin{equation}
    \sup_{z\in \mathfrak{K}}\left\Vert \Psi_{K^{\mathrm{ref}}}(\eta,z)-\psi_{K^{\mathrm{ref}}}(\cdot ,z)-D_\eta\Psi_{K^{\mathrm{ref}}}(0,z)[\eta]\right\Vert
    _{\infty ,[0,T]}\leq C_{\psi ,\mathfrak{K}}\Vert \eta\Vert _{L^{1}(0,T)}^{2}
    \label{eq:uniform_quadratic_solution}
\end{equation}
for every $\eta\in B_{X}(0,r_{\psi, \mathfrak{K}})$.

\item[(ii)] For $z\in \{0,1\}$,
\begin{equation}
    \Psi_{K^{\mathrm{ref}}}(\eta,z)\equiv 0  \label{eq:extended_solution_endpoints}
\end{equation}%
for every $\eta\in X$ such that $(\eta,z)\in \mathcal{O}_{K^{\mathrm{ref}}\times \mathcal{Z}}$.
\end{itemize}
\end{proposition}

\begin{proof}
    (i) Let $\mathfrak{K}\subset \mathcal{Z}$ be compact. Since $\mathcal O_{K^{\mathrm{ref}}\times\mathcal{Z}}$ is open and contains $\{0\}\times\mathfrak{K}$, for every $z\in\mathfrak K$ there exist a number $\rho_{z}>0$ and an open neighborhood $\mathcal V_{z}\subset\mathbb C$ of $z$ such that
	\begin{equation*}
		B_X(0,\rho_{z})\times\mathcal V_{z}
		\subset
		\mathcal O_{K^{\mathrm{ref}}\times\mathcal{Z}}.
	\end{equation*}
	Since $\Psi_{K^{\mathrm{ref}}}$ is real-analytic, the map $(\eta,z)\longmapsto D_\eta^2\Psi_{K^{\mathrm{ref}}}(\eta,z)$ is continuous in the operator norm $\mathscr L_{\mathbb R}^{2}(X,X;Y):=\mathscr L_{\mathbb R}(X,\mathscr L_{\mathbb R}(X,Y))$. After decreasing $\rho_{z}$ and $\mathcal V_{z}$ if necessary, there exists $M_{z}<\infty$ such that
	\begin{equation*}
        \sup_{B_{X}(0,\rho_{z})\times \mathcal{V}_{z}}\Vert D_{\eta}^{2}\Psi_{K^{\mathrm{ref}}}(\eta,z)\Vert_{\mathscr L_{\mathbb R}^{2}(X,X;Y)}\leq M_{z}.
    \end{equation*}
	On the other hand, the family $\{\mathcal V_{z}:z\in\mathfrak K\}$ covers $\mathfrak K$. By compactness, there are $z_1,\ldots,z_N\in\mathfrak K$ such that $\mathfrak K\subset\bigcup_{j=1}^N\mathcal V_{z_j}$.
	Set
	\begin{equation*}
		r_{\psi, \mathfrak K}
		:=
		\min_{1\leq j\leq N}\rho_{z_j}>0,
		\qquad
		M_{\mathfrak K}
		:=
		\max_{1\leq j\leq N}M_{z_j}<\infty.
	\end{equation*}
	Then
	\begin{equation*}
		B_X(0,r_{\psi, \mathfrak K})\times\mathfrak K
		\subset\mathcal O_{K^{\mathrm{ref}}\times\mathcal{Z}}
	\end{equation*}
	and
	\begin{equation}
		\sup_{\substack{
			z\in\mathfrak K\\
			\eta\in B_X(0,r_{\psi, \mathfrak K})
		}}
		\|D_\eta^2\Psi_{K^{\mathrm{ref}}}(\eta,z)\|_{\mathscr L_{\mathbb R}^{2}(X,X;Y)}
		\leq
		M_{\mathfrak K}.
		\label{eq:uniform_second_derivative_solution}
	\end{equation}

	For fixed $z\in\mathfrak K$ and $\eta\in B_X(0,r_{\psi, \mathfrak K})$, the segment $\{(\theta \eta,z):\theta\in[0,1]\}$ is contained in $\mathcal O_{K^{\mathrm{ref}}\times\mathcal{Z}}$. Taylor formula with integral remainder for mappings between Banach spaces \cite[XIII,\S 6 ]{lang1993} gives
	\begin{align*}
		\Psi_{K^{\mathrm{ref}}}(\eta,z)
		&=
		\Psi_{K^{\mathrm{ref}}}(0,z)
		+
		D_\eta\Psi_{K^{\mathrm{ref}}}(0,z)[\eta]
		+
		\int_0^1
		(1-\theta)
		D_\eta^2\Psi_{K^{\mathrm{ref}}}(\theta \eta,z)[\eta,\eta]\,d\theta.
	\end{align*}
	Using \eqref{eq:extended_solution_at_reference} and \eqref{eq:uniform_second_derivative_solution}, we obtain
	\begin{equation*}
		\left\|
			\Psi_{K^{\mathrm{ref}}}(\eta,z)
			-
			\psi_{K^{\mathrm{ref}}}(\cdot,z)
			-
			D_\eta\Psi_{K^{\mathrm{ref}}}(0,z)[\eta]
		\right\|_{\infty,[0,T]}
		\leq
		\frac{M_{\mathfrak K}}{2}
		\|\eta\|_{L^1(0,T)}^2.
	\end{equation*}
	Taking the supremum over $z\in\mathfrak K$ proves \eqref{eq:uniform_quadratic_solution} with $C_{\psi,\mathfrak K}:= \frac{M_{\mathfrak K}}{2}$.

	(ii) Let $z\in\{0,1\}$ and $(\eta,z)\in\mathcal O_{K^{\mathrm{ref}}\times\mathcal{Z}}$. Since $F(z,0)=0$, the zero function is a solution of \eqref{eq:extended_riccati}. Both $K^{\mathrm{ref}}$ and $\eta$ belong to $L^1_{\mathbb R}(0,T)$, so $K^{\mathrm{ref}}+\eta\in L^1_{\mathbb R}(0,T)$. Lemma \ref{lem:riccati_uniqueness}, applied with $J=K^{\mathrm{ref}}+\eta$, therefore implies that this solution is unique. Hence $\Psi_{K^{\mathrm{ref}}}(\eta,z)\equiv0$, proving
	\eqref{eq:extended_solution_endpoints}.
\end{proof}

\subsection{Adjoint representation of the kernel derivative}
\label{subsec:linearized_dual}

On the open set $\mathcal O_{K^{\mathrm{ref}}\times\mathcal Z}$
obtained in Theorem~\ref{thm:local_analytic_solution}, define
\begin{equation}
	\widetilde{\mathcal L}_{K^{\mathrm{ref}}}(\eta, z)
	:=
	\int_0^T
	F\bigl(z,\Psi_{K^{\mathrm{ref}}}(\eta,z)(s)\bigr)
	g(T-s)\,ds.
	\label{eq:extended_riccati_functional}
\end{equation}
This defines a real-analytic map $\widetilde{\mathcal L}_{K^{\mathrm{ref}}}: \mathcal O_{K^{\mathrm{ref}}\times\mathcal{Z}}\longrightarrow\mathbb C$.
Indeed, by Theorem~\ref{thm:local_analytic_solution}, $N(\eta,z):=\bigl(\Psi_{K^{\mathrm{ref}}}(\eta,z),z\bigr)$ defines a real-analytic map $N:\mathcal O_{K^{\mathrm{ref}}\times\mathcal Z}\longrightarrow Y\times\mathbb C$. 
Moreover, $G:Y\times\mathbb C\longrightarrow Y$, defined by $G(\varphi,z)(s):=F(z,\varphi(s))$, is real-analytic, since the Nemytskii map $(z,\varphi)\mapsto F(z,\varphi)$ is a polynomial map. 
Finally, the map $\beta:Y\longrightarrow\mathbb C$, defined by
$\beta(f):=\int_0^T f(s)g(T-s)\,ds$ for $f\in Y$,
is bounded and linear, since Assumption~\ref{ass:model}(iii) gives
\begin{equation*}
    |\beta(f)|
    = 
    \left\vert \int_{0}^{T}f(s) g(T-s)ds\right\vert
    \le
    T\|g\|_{\infty,[0,T]}
    \|f\|_{\infty,[0,T]}.
\end{equation*}
Hence, $\beta$ is real-analytic and $\widetilde{\mathcal L}_{K^{\mathrm{ref}}}=\beta\circ G\circ N$.
Therefore, by \cite[Theorem 4.5.7]{buffoni_toland}, $\widetilde{\mathcal L}_{K^{\mathrm{ref}}}:\mathcal O_{K^{\mathrm{ref}}\times\mathcal Z}\to\mathbb C$ is real-analytic.
This proves the following proposition.

\begin{proposition}
\label{prop:extended_functional_analytic}
 Suppose that Assumptions~\ref{ass:model}(ii)--(iii) and \ref{ass:reference_riccati} hold. The map $\widetilde{\mathcal{L}}_{K^{\mathrm{ref}}}$ defined by %
\eqref{eq:extended_riccati_functional} is well defined and real-analytic on $\mathcal{O}_{K^{\mathrm{ref}}\times \mathcal{Z}}$.
\end{proposition}

If $K^{\mathrm{ref}}\in \mathscr K_T$, then, for every
$z\in\mathcal Z$,
\begin{equation}
    \mathcal L_T(K^{\mathrm{ref}}, z)
	=
    \widetilde{\mathcal L}_{K^{\mathrm{ref}}}(0, z).
	\label{eq:extended_functional_at_reference}
\end{equation}
Moreover, whenever $z\in\mathcal Z$,
$(\eta,z)\in\mathcal O_{K^{\mathrm{ref}}\times\mathcal Z}$
and $K^{\mathrm{ref}}+\eta\in \mathscr K_T$, one has
\begin{equation}
    \mathcal L_T(K^{\mathrm{ref}}+\eta,z)
	=
	\widetilde{\mathcal L}_{K^{\mathrm{ref}}}(\eta,z).
	\label{eq:extended_functional_coincidence}
\end{equation}
These identities follow from
\eqref{eq:extended_solution_at_reference} and
\eqref{eq:extended_solution_coincidence}, respectively.

We next compute the partial Fr\'{e}chet derivative with respect to the
kernel variable. For $z\in\mathbb C$ and
$\varphi,w\in Y$, the pointwise identity
\begin{equation*}
	F(z,\varphi+w)
	-
	F(z,\varphi)
	-
	\partial_xF(z,\varphi)w
	=
	\frac{\nu^2}{2}w^2
\end{equation*}
shows that the derivative of
$\varphi\mapsto G(\varphi, z)$ is the bounded linear operator $w\longmapsto\partial_xF(z,\varphi)w$, where multiplication is understood pointwise. Hence, by the chain rule for Fr\'{e}chet derivatives, for every $(\eta,z)\in\mathcal O_{K^{\mathrm{ref}}\times\mathcal{Z}}$ and $h\in X$,
\begin{equation}
	D_\eta\widetilde{\mathcal L}_{K^{\mathrm{ref}}}(\eta,z)[h]
	=
	\int_0^T
	\partial_xF
	\bigl(z,\Psi_{K^{\mathrm{ref}}}(\eta,z)(s)\bigr)
	D_\eta\Psi_{K^{\mathrm{ref}}}(\eta,z)[h](s)
	g(T-s)\,ds.
	\label{eq:general_first_variation}
\end{equation}

Fix $z\in\mathcal{Z}$. Recall that
\begin{equation*}
	q_{K^{\mathrm{ref}},z}(s)
	:=
	\partial_xF\bigl(z,\psi_{K^{\mathrm{ref}}}(s,z)\bigr)
	=
	(\rho\nu z-\lambda)
	+
	\nu^2\psi_{K^{\mathrm{ref}}}(s,z),
	\qquad s\in[0,T],
\end{equation*}
and let $\mathcal{A}_{K^{\mathrm{ref}},z}$ be the linearized Volterra operator defined in \eqref{eq:linearized_operator}. Evaluating
\eqref{eq:general_first_variation} at $\eta=0$ and using
\eqref{eq:derivative_solution_map}, assuming Assumption~\ref{ass:reference_riccati}, we obtain, for every $h\in X$,
\begin{align}
	D_\eta\widetilde{\mathcal L}_{K^{\mathrm{ref}}}(0,z)[h]
	&=
	\int_0^T
	q_{K^{\mathrm{ref}},z}(s)
	D_\eta\Psi_{K^{\mathrm{ref}}}(0,z)[h](s)
	g(T-s)\,ds
	\nonumber\\
	&=
	\int_0^T
	\left[
		(I-\mathcal A_{K^{\mathrm{ref}},z})^{-1}
		\left(
			h*
			F\bigl(z,\psi_{K^{\mathrm{ref}}}(\cdot,z)\bigr)
		\right)
	\right](s)
	G_{K^{\mathrm{ref}},z}(s)\,ds,
	\label{eq:first_variation_before_transposition}
\end{align}
where
\begin{equation}
	G_{K^{\mathrm{ref}},z}(s)
	:=
	q_{K^{\mathrm{ref}},z}(s)g(T-s),
	\qquad s\in[0,T].
	\label{eq:dual_source}
\end{equation}
Since
$q_{K^{\mathrm{ref}},z}\in C([0,T];\mathbb C)$ and
$g\in C([0,T];\mathbb{R}_{+})$, one has $G_{K^{\mathrm{ref}},z}\in Y$.

Furthermore, $F\bigl(z,\psi_{K^{\mathrm{ref}}}(\cdot,z)\bigr)\in Y$, and therefore $h\ast F\bigl(z,\psi_{K^{\mathrm{ref}}}(\cdot,z)\bigr)\in Y$. Indeed, the translation argument used in the proof of
Proposition~\ref{prop:linearized_invertibility} shows that the
Volterra convolution of an $L^{1}$ kernel with a continuous function is
continuous, while
\begin{equation}
	\left\|
		h\ast
		F\bigl(z,\psi_{K^{\mathrm{ref}}}(\cdot,z)\bigr)
	\right\|_{\infty,[0,T]}
	\leq
	\|h\|_{L^1(0,T)}
	\left\|
		F\bigl(z,\psi_{K^{\mathrm{ref}}}(\cdot,z)\bigr)
	\right\|_{\infty,[0,T]}.
	\label{eq:linearized_source_bound}
\end{equation}
Thus, every term in \eqref{eq:first_variation_before_transposition} is well defined. As a partial Fr\'{e}chet derivative,
\begin{equation*}
	D_\eta\widetilde{\mathcal L}_{K^{\mathrm{ref}}}(0,z)
	\in
	\mathscr L_{\mathbb R}(X,\mathbb C).
\end{equation*}

To derive a dual representation of this functional, we introduce the
continuous complex-bilinear form
\begin{equation}
	\langle\varphi,w\rangle
	:=
	\int_0^T\varphi(t)w(t)\,dt,
	\qquad
	\varphi,w\in Y.
	\label{eq:bilinear_pairing}
\end{equation}
It satisfies
\begin{equation}
	|\langle\varphi,w\rangle|
	\leq
	T
	\|\varphi\|_{\infty,[0,T]}
	\|w\|_{\infty,[0,T]}.
	\label{eq:bilinear_pairing_bound}
\end{equation}
No complex conjugation is involved, so
\eqref{eq:bilinear_pairing} is not a Hermitian inner product; it is an
auxiliary bilinear form used to transpose the linearized Volterra
operator.

With this notation,
\begin{equation}
	D_\eta\widetilde{\mathcal L}_{K^{\mathrm{ref}}}(0,z)[h]
	=
	\left\langle
		(I-\mathcal A_{K^{\mathrm{ref}},z})^{-1}
		\left(
			h\ast
			F\bigl(z,\psi_{K^{\mathrm{ref}}}(\cdot,z)\bigr)
		\right),
		G_{K^{\mathrm{ref}},z}
	\right\rangle.
	\label{eq:first_variation_pairing}
\end{equation}

We next define the transpose of $\mathcal A_{K^{\mathrm{ref}},z}$ with respect to \eqref{eq:bilinear_pairing} and show that it is a backward Volterra operator. The resulting transposition identity will transfer the
inverse operator in \eqref{eq:first_variation_pairing} from the forward linearized Riccati response to the source $G_{K^{\mathrm{ref}},z}$.

\begin{proposition}
	\label{prop:transpose_operator}
	Suppose that Assumptions~\ref{ass:model}(ii)--(iii) and \ref{ass:reference_riccati} hold. Fix $z\in\mathcal{Z}$. Define
	$\mathcal A_{K^{\mathrm{ref}},z}^{\top}:Y\to Y$ by
	\begin{equation}
		(\mathcal A_{K^{\mathrm{ref}},z}^{\top}w)(s)
		:=
		q_{K^{\mathrm{ref}},z}(s)
		\int_s^T K^{\mathrm{ref}}(t-s)w(t)\,dt,
		\qquad s\in[0,T].
		\label{eq:transpose_operator}
	\end{equation}

    The following statements hold:
    \begin{itemize}
	\item[(i)] $\mathcal A_{K^{\mathrm{ref}},z}^{\top}$ is a bounded $\mathbb C$-linear operator satisfying
	\begin{equation}
		\|\mathcal A_{K^{\mathrm{ref}},z}^{\top}\|_{\mathscr L_{\mathbb{C}}(Y,Y)}
		\leq
		\|K^{\mathrm{ref}}\|_{L^1(0,T)}
		\|q_{K^{\mathrm{ref}},z}\|_{\infty,[0,T]},
		\label{eq:transpose_operator_bound}
	\end{equation}
	and
	\begin{equation}
		\langle\mathcal A_{K^{\mathrm{ref}},z}\varphi,w\rangle
		=
		\langle\varphi,\mathcal A_{K^{\mathrm{ref}},z}^{\top}w\rangle
		\label{eq:transpose_identity}
	\end{equation}
	for every $\varphi,w\in Y$.
    \item[(ii)] 	The operator $I-\mathcal A_{K^{\mathrm{ref}},z}^{\top}$ is boundedly
	invertible on $Y$. Moreover,
	\begin{equation}
		\left\langle
			(I-\mathcal A_{K^{\mathrm{ref}},z})^{-1}\varphi,w
		\right\rangle
		=
		\left\langle
			\varphi,
			(I-\mathcal A_{K^{\mathrm{ref}},z}^{\top})^{-1}w
		\right\rangle
		\label{eq:inverse_transposition}
	\end{equation}
	for every $\varphi,w\in Y$.
	\end{itemize}
\end{proposition}

\begin{proof}
	(i) First, we prove that $\mathcal A_{K^{\mathrm{ref}},z}^{\top}$ is well defined. Let $w\in Y$, and denote by $\overline w$ its extension by zero
	outside $[0,T]$. After the change of variables $u=t-s$,
	\begin{equation}
		\int_s^T K^{\mathrm{ref}}(t-s)w(t)\,dt
		=
		\int_0^T K^{\mathrm{ref}}(u)\overline w(s+u)\,du.
		\label{eq:backward_integral_extension}
	\end{equation}
	The integral is well defined because $K^{\mathrm{ref}}\in L^1_{\mathbb R}(0,T)$ and $\overline w$ is bounded.

	Suppose that $s_n\to s$ in $[0,T]$. Then
	\begin{equation*}
		\overline w(s_n+u)
		\longrightarrow
		\overline w(s+u)
	\end{equation*}
	for almost every $u\in(0,T)$; the possible exceptional values are
	those for which $s+u$ is an endpoint of $[0,T]$. Moreover,
	\begin{equation*}
		|K^{\mathrm{ref}}(u)\overline w(s_n+u)|
		\leq
		|K^{\mathrm{ref}}(u)|\|w\|_{\infty,[0,T]},
	\end{equation*}
	and the right-hand side is integrable. Dominated convergence \cite[Theorem 4.2]{brezis2011functional} applied to \eqref{eq:backward_integral_extension}, together with
	the continuity of $q_{K^{\mathrm{ref}},z}$, therefore shows that
	$\mathcal A_{K^{\mathrm{ref}},z}^{\top}w\in Y$.

	The $\mathbb C$-linearity is immediate, and
	\begin{equation*}
		|(\mathcal A_{K^{\mathrm{ref}},z}^{\top}w)(s)|
		\leq
		|q_{K^{\mathrm{ref}},z}(s)|
		\|w\|_{\infty,[0,T]}
		\int_s^T |K^{\mathrm{ref}}(t-s)|\,dt
		\leq
		\|q_{K^{\mathrm{ref}},z}\|_{\infty,[0,T]}
		\|K^{\mathrm{ref}}\|_{L^1(0,T)}
		\|w\|_{\infty,[0,T]}.
	\end{equation*}
	Taking the supremum over $s\in[0,T]$ proves
	\eqref{eq:transpose_operator_bound}.

	Let $\varphi,w\in Y$. The function $(s,t)\longmapsto K^{\mathrm{ref}}(t-s)q_{K^{\mathrm{ref}},z}(s)\varphi(s)w(t)$ is absolutely integrable on the Volterra triangle $\Delta_T:=\{(s,t)\in[0,T]^2:0\leq s\leq t\leq T\}$, since
	\begin{align*}
		&\int_0^T\int_0^t
		\left|
			K^{\mathrm{ref}}(t-s)q_{K^{\mathrm{ref}},z}(s)\varphi(s)w(t)
		\right|\,ds\,dt
		\\
		&\qquad\leq
		T\|K^{\mathrm{ref}}\|_{L^1(0,T)}
		\|q_{K^{\mathrm{ref}},z}\|_{\infty,[0,T]}
		\|\varphi\|_{\infty,[0,T]}
		\|w\|_{\infty,[0,T]}.
	\end{align*}
	Fubini theorem \cite[Theorem 4.5]{brezis2011functional} therefore yields
	\begin{align*}
		\langle\mathcal A_{K^{\mathrm{ref}},z}\varphi,w\rangle
		&=
		\int_0^T\int_0^t
		K^{\mathrm{ref}}(t-s)q_{K^{\mathrm{ref}},z}(s)\varphi(s)w(t)\,ds\,dt
		\\
		&=
		\int_0^T
		\varphi(s)q_{K^{\mathrm{ref}},z}(s)
		\int_s^T K^{\mathrm{ref}}(t-s)w(t)\,dt\,ds
		\\
		&=
		\langle\varphi,\mathcal A_{K^{\mathrm{ref}},z}^{\top}w\rangle,
	\end{align*}
	which proves \eqref{eq:transpose_identity}.

	(ii) To establish invertibility, define the time-reversal operator $\mathcal J:Y\to Y$ by $(\mathcal Jw)(t):=w(T-t)$, $t\in[0,T]$. It is a bounded isometric involution, since $\mathcal J^2=I$ and $\|\mathcal Jw\|_{\infty,[0,T]}=\|w\|_{\infty,[0,T]}$. A direct change of variables gives
	\begin{equation}
		(\mathcal J\mathcal A_{K^{\mathrm{ref}},z}^{\top}\mathcal Jw)(\tau)
		=
		q_{K^{\mathrm{ref}},z}(T-\tau)
		\int_0^\tau K^{\mathrm{ref}}(\tau-\sigma)w(\sigma)\,d\sigma.
		\label{eq:time_reversed_transpose}
	\end{equation}
	Thus, $\mathcal J\mathcal A_{K^{\mathrm{ref}},z}^{\top}\mathcal J$ is a forward Volterra operator with kernel
	\begin{equation*}
		b_{K^{\mathrm{ref}},z}(\tau,\sigma)
		:=
		\overline{K^{\mathrm{ref}}}(\tau-\sigma)
		q_{K^{\mathrm{ref}},z}(T-\tau),
		\qquad
		(\tau,\sigma)\in[0,T]^2,
	\end{equation*}
	where $\overline{K^{\mathrm{ref}}}$ denotes the extension of $K^{\mathrm{ref}}$ by zero outside $(0,T)$.

	For $\tau,\tau_0\in[0,T]$,
	\begin{align*}
		&
		\|b_{K^{\mathrm{ref}},z}(\tau,\cdot)
		-
		b_{K^{\mathrm{ref}},z}(\tau_0,\cdot)\|_{L^1(0,T)}
		\\
		&\quad\leq
		|q_{K^{\mathrm{ref}},z}(T-\tau)|
		\|
			\overline{K^{\mathrm{ref}}}(\tau-\cdot)
			-
			\overline{K^{\mathrm{ref}}}(\tau_0-\cdot)
		\|_{L^1(\mathbb R)}
		\\
		&\qquad+
		|q_{K^{\mathrm{ref}},z}(T-\tau)-q_{K^{\mathrm{ref}},z}(T-\tau_0)|
		\|K^\mathrm{ref}\|_{L^1(0,T)}.
	\end{align*}
	The first term tends to zero as $\tau\to\tau_0$ by the strong
	continuity of translations in $L^1(\mathbb R)$, and the second
	one tends to zero by the continuity of $q_{K^{\mathrm{ref}},z}$. Furthermore,
	\begin{equation*}
		\sup_{\tau\in[0,T]}
		\|b_{K^{\mathrm{ref}},z}(\tau,\cdot)\|_{L^1(0,T)}
		\leq
		\|q_{K^{\mathrm{ref}},z}\|_{\infty,[0,T]}
		\|K^{\mathrm{ref}}\|_{L^1(0,T)}.
	\end{equation*}
    Since $[0,T]$ is compact, this $L^1$-continuity is uniform. Together with the preceding uniform bound, the uniform $L^1$-continuity shows that $b_{K^{\mathrm{ref}},z}$, and hence also $-b_{K^{\mathrm{ref}},z}$, is a Volterra kernel of bounded uniformly continuous type. By \cite[Theorem~9.5.5(ii)]{gripenberg}, $-b_{K^{\mathrm{ref}},z}$ has a resolvent of bounded uniformly continuous type on $[0,T]$, which is in particular of type $L^\infty$. Since every $f\in Y=C([0,T];\mathbb C)$ is bounded and uniformly continuous, \cite[Theorem~9.5.4]{gripenberg} yields a unique $\xi\in Y$ satisfying $\bigl(I-\mathcal J\mathcal A_{K^{\mathrm{ref}},z}^{\top}\mathcal J\bigr)\xi=f$. Hence, $I-\mathcal J\mathcal A_{K^{\mathrm{ref}},z}^{\top}\mathcal J$ is a bounded linear bijection on the Banach space $Y$, so its inverse is bounded by the bounded inverse theorem.

	Since
	\begin{equation*}
		I-\mathcal J\mathcal A_{K^{\mathrm{ref}},z}^{\top}\mathcal J
		=
		\mathcal J
		(I-\mathcal A_{K^{\mathrm{ref}},z}^{\top})
		\mathcal J,
	\end{equation*}
	and $\mathcal J^{-1}=\mathcal J$, it follows that
	$I-\mathcal A_{K^{\mathrm{ref}},z}^{\top}$ is boundedly invertible, with
	\begin{equation*}
		(I-\mathcal A_{K^{\mathrm{ref}},z}^{\top})^{-1}
		=
		\mathcal J
		(I-\mathcal J\mathcal A_{K^{\mathrm{ref}},z}^{\top}\mathcal J)^{-1}
		\mathcal J.
	\end{equation*}

	Finally, let
	\begin{equation*}
		u:=(I-\mathcal A_{K^{\mathrm{ref}},z})^{-1}\varphi,
		\qquad
		v:=(I-\mathcal A_{K^{\mathrm{ref}},z}^{\top})^{-1}w.
	\end{equation*}
	Then
	\begin{equation*}
		(I-\mathcal A_{K^{\mathrm{ref}},z})u=\varphi,
		\qquad
		(I-\mathcal A_{K^{\mathrm{ref}},z}^{\top})v=w.
	\end{equation*}
	Using \eqref{eq:transpose_identity},
	\begin{align*}
		\langle u,w\rangle
		&=
		\langle u,(I-\mathcal A_{K^{\mathrm{ref}},z}^{\top})v\rangle
		\\
		&=
		\langle u,v\rangle
		-
		\langle u,\mathcal A_{K^{\mathrm{ref}},z}^{\top}v\rangle
		\\
		&=
		\langle u,v\rangle
		-
		\langle\mathcal A_{K^{\mathrm{ref}},z}u,v\rangle
		\\
		&=
		\langle(I-\mathcal A_{K^{\mathrm{ref}},z})u,v\rangle
		=
		\langle\varphi,v\rangle.
	\end{align*}
	This is precisely \eqref{eq:inverse_transposition}.
\end{proof}

\begin{definition}
	\label{def:dual_state}
	For fixed $z\in\mathcal Z$, define the adjoint state associated with
	$K^{\mathrm{ref}}$ and $z$ by
	\begin{equation}
		Y_{K^{\mathrm{ref}},z}
		:=
		(I-\mathcal A_{K^{\mathrm{ref}},z}^{\top})^{-1}G_{K^{\mathrm{ref}},z}
		\in Y.
		\label{eq:dual_state}
	\end{equation}
\end{definition}

By Proposition~\ref{prop:transpose_operator}, $I-\mathcal A_{K^{\mathrm{ref}},z}^{\top}$ is boundedly invertible on $Y$.
Consequently, $Y_{K^{\mathrm{ref}},z}$ is the unique function in $Y$ satisfying
\begin{equation}
	Y_{K^{\mathrm{ref}},z}(s)
	-
	q_{K^{\mathrm{ref}},z}(s)
	\int_s^T
	K^{\mathrm{ref}}(t-s)Y_{K^{\mathrm{ref}},z}(t)\,dt
	=
	q_{K^{\mathrm{ref}},z}(s)g(T-s),
	\qquad s\in[0,T].
	\label{eq:dual_state_backward}
\end{equation}
Moreover,
\begin{equation}
	\|Y_{K^{\mathrm{ref}},z}\|_{\infty,[0,T]}
	\leq
	\bigl\|
		(I-\mathcal A_{K^{\mathrm{ref}},z}^{\top})^{-1}
	\bigr\|_{\mathscr L_{\mathbb C}(Y,Y)}
	\,
	\|G_{K^{\mathrm{ref}},z}\|_{\infty,[0,T]}.
	\label{eq:dual_state_bound}
\end{equation}

For later use, for each $z\in\mathcal Z$, set
\begin{equation}
	f_{K^{\mathrm{ref}},z}(s)
	:=
	F\bigl(z,\psi_{K^{\mathrm{ref}}}(s,z)\bigr),
	\qquad s\in[0,T].
	\label{eq:reference_riccati_source}
\end{equation}
Since $\psi_{K^{\mathrm{ref}}}(\cdot,z)\in Y$ and $F$ is continuous,
we have $f_{K^{\mathrm{ref}},z}\in Y$.

\begin{lemma}
	\label{prop:sensitivity_kernel}
	Suppose that Assumptions~\ref{ass:model}(ii)--(iii) and \ref{ass:reference_riccati} hold. For fixed $z\in\mathcal Z$, define
	\begin{equation}
		B_{K^{\mathrm{ref}}}(u,z)
		:=
		\int_0^{T-u}
		F\bigl(z,\psi_{K^{\mathrm{ref}}}(\sigma,z)\bigr)
		Y_{K^{\mathrm{ref}},z}(\sigma+u)\,d\sigma,
		\qquad u\in[0,T].
		\label{eq:sensitivity_kernel}
	\end{equation}
	Then
	\begin{equation*}
		B_{K^{\mathrm{ref}}}(\cdot,z)\in C([0,T];\mathbb C),
		\qquad
		B_{K^{\mathrm{ref}}}(T,z)=0,
	\end{equation*}
	and
	\begin{equation}
		\|B_{K^{\mathrm{ref}}}(\cdot,z)\|_{\infty,[0,T]}
		\leq
		T
		\bigl\|
			F\bigl(z,\psi_{K^{\mathrm{ref}}}(\cdot,z)\bigr)
		\bigr\|_{\infty,[0,T]}
		\|Y_{K^{\mathrm{ref}},z}\|_{\infty,[0,T]}.
		\label{eq:sensitivity_kernel_bound}
	\end{equation}
\end{lemma}

\begin{proof}
	By \eqref{eq:reference_riccati_source}, for every $u\in[0,T]$, definition
	\eqref{eq:sensitivity_kernel} gives
	\begin{align*}
		|B_{K^{\mathrm{ref}}}(u,z)|
		&\leq
		\int_0^{T-u}
		|f_{K^{\mathrm{ref}},z}(\sigma)|
		|Y_{K^{\mathrm{ref}},z}(\sigma+u)|\,d\sigma
		\\
		&\leq
		(T-u)
		\|f_{K^{\mathrm{ref}},z}\|_{\infty,[0,T]}
		\|Y_{K^{\mathrm{ref}},z}\|_{\infty,[0,T]}.
	\end{align*}
	Taking the supremum over $u\in[0,T]$ proves
	\eqref{eq:sensitivity_kernel_bound}.

	It remains to establish continuity. Let
	$\overline Y_{K^{\mathrm{ref}},z}:\mathbb R\to\mathbb C$ denote the extension of $Y_{K^{\mathrm{ref}},z}$ by zero outside $[0,T]$. For $u\in[0,T]$,
	\begin{equation}
		B_{K^{\mathrm{ref}}}(u,z)
		=
		\int_0^T
		f_{K^{\mathrm{ref}},z}(\sigma)
		\overline{Y}_{K^{\mathrm{ref}},z}(\sigma+u)\,d\sigma.
		\label{eq:sensitivity_kernel_extension}
	\end{equation}
	If $u_n\to u$ in $[0,T]$, then
	\begin{equation*}
		\overline{Y}_{K^{\mathrm{ref}},z}(\sigma+u_n)
		\longrightarrow
		\overline Y_{K^{\mathrm{ref}},z}(\sigma+u)
	\end{equation*}
	for almost every $\sigma\in(0,T)$; the only possible exceptional
	points correspond to $\sigma+u\in\{0,T\}$. Moreover,
	\begin{equation*}
		\left|
			f_{K^{\mathrm{ref}},z}(\sigma)
			\overline{Y}_{K^{\mathrm{ref}},z}(\sigma+u_n)
		\right|
		\leq
		\|f_{K^{\mathrm{ref}},z}\|_{\infty,[0,T]}
		\|Y_{K^{\mathrm{ref}},z}\|_{\infty,[0,T]},
	\end{equation*}
	and the right-hand side is integrable on $(0,T)$. Dominated
	convergence applied to
	\eqref{eq:sensitivity_kernel_extension} therefore gives
	\begin{equation*}
		B_{K^{\mathrm{ref}}}(u_n,z)\longrightarrow B_{K^{\mathrm{ref}}}(u,z).
	\end{equation*}
	Hence $B_{K^{\mathrm{ref}}}(\cdot,z)\in C([0,T];\mathbb C)$. Finally, \eqref{eq:sensitivity_kernel} gives $B_{K^{\mathrm{ref}}}(T,z)=0$, since the integration interval degenerates to a single point.
\end{proof}

\begin{theorem}
	\label{thm:frechet_diff}
	Under Assumptions~\ref{ass:model}(ii)--(iii) and \ref{ass:reference_riccati}, define, for every $z\in\mathcal Z$,
	\begin{equation}
		\Lambda_{K^{\mathrm{ref}},z}[h]
		:=
		\int_0^T B_{K^{\mathrm{ref}}}(u,z)h(u)\,du,
		\qquad h\in X.
		\label{eq:sensitivity_functional}
	\end{equation}
	Then
	\begin{equation}
		D_\eta\widetilde{\mathcal L}_{K^{\mathrm{ref}}}(0,z)
		=
		\Lambda_{K^{\mathrm{ref}},z}.
		\label{eq:frechet_derivative_identification}
	\end{equation}
\end{theorem}

\begin{proof}
    By \eqref{eq:first_variation_pairing},
	\eqref{eq:inverse_transposition}, and
	\eqref{eq:dual_state}, for every $h\in X $,
    \begin{align}
        D_{\eta}\widetilde{\mathcal{L}}_{K^{\mathrm{ref}}}(0,z)[h]& =\left\langle (I-\mathcal{A}%
        _{K^{\mathrm{ref}},z})^{-1}(h\ast f_{K^{\mathrm{ref}},z}),G_{K^{\mathrm{ref}},z}\right\rangle  \notag \\
        & =\left\langle h\ast f_{K^{\mathrm{ref}},z},(I-\mathcal{A}_{K^{\mathrm{ref}},z}^{\top
        })^{-1}G_{K^{\mathrm{ref}},z}\right\rangle  \notag \\
        & =\int_{0}^{T}(h\ast f_{K^{\mathrm{ref}},z})(t)Y_{K^{\mathrm{ref}},z}(t)\,dt.
        \label{eq:first_variation_dual_state}
    \end{align}
    Expanding the Volterra convolution gives
    \begin{equation*}
        \int_{0}^{T}\int_{0}^{t}h(t-\sigma )f_{K^{\mathrm{ref}},z}(\sigma )Y_{K^{\mathrm{ref}},z}(t)\,d\sigma \,dt.
    \end{equation*}
    This integrand is absolutely integrable. Indeed,
    \begin{align*}
        & \int_{0}^{T}\int_{0}^{t}|h(t-\sigma )||f_{K^{\mathrm{ref}},z}(\sigma
        )||Y_{K^{\mathrm{ref}},z}(t)|\,d\sigma \,dt \\
        & \qquad \leq \Vert f_{K^{\mathrm{ref}},z}\Vert _{\infty ,[0,T]}\Vert Y_{K^{\mathrm{ref}},z}\Vert _{\infty
        ,[0,T]}\int_{0}^{T}\int_{0}^{t}|h(t-\sigma )|\,d\sigma \,dt \\
        & \qquad =\Vert f_{K^{\mathrm{ref}},z}\Vert _{\infty ,[0,T]}\Vert Y_{K^{\mathrm{ref}},z}\Vert _{\infty
        ,[0,T]}\int_{0}^{T}(T-u)|h(u)|\,du \\
        & \qquad \leq T\Vert f_{K^{\mathrm{ref}},z}\Vert _{\infty ,[0,T]}\Vert Y_{K^{\mathrm{ref}},z}\Vert
        _{\infty ,[0,T]}\Vert h\Vert _{L^{1}(0,T)}.
    \end{align*}%
    Fubini theorem is therefore applicable. Using the change of variables $u=t-\sigma$, $t=\sigma +u$, which maps the Volterra triangle onto $\{(\sigma ,u)\in \lbrack 0,T]^{2}:0\leq u\leq T,\ 0\leq \sigma \leq T-u\}$, we obtain
    \begin{align*}
        D_{\eta}\widetilde{\mathcal{L}}_{K^{\mathrm{ref}}}(0,z)[h]&
        =\int_{0}^{T}h(u)\int_{0}^{T-u}f_{K^{\mathrm{ref}},z}(\sigma )Y_{K^{\mathrm{ref}},z}(\sigma +u)\,d\sigma
        \,du \\
        & =\int_{0}^{T}B_{K^{\mathrm{ref}}}(u,z)h(u)\,du
    \end{align*}%
    and therefore
    \begin{equation*}
        D_{\eta}\widetilde{\mathcal{L}}_{K^{\mathrm{ref}}}(0,z)[h]=\Lambda _{K^{\mathrm{ref}},z}[h].
    \end{equation*}%
    $\Lambda _{K^{\mathrm{ref}},z}\in \mathscr {L}_{\mathbb{R}}(X,\mathbb{C})$, since it is
   linear and satisfies
    \begin{equation}
        |\Lambda _{K^{\mathrm{ref}},z}[h]|\leq \Vert B_{K^{\mathrm{ref}}}(\cdot ,z)\Vert _{\infty ,[0,T]}\Vert
        h\Vert _{L^{1}(0,T)},\qquad h\in X,  \label{eq:sensitivity_functional_bound}
    \end{equation}
    which proves \eqref{eq:frechet_derivative_identification}.
\end{proof}

The first-order Taylor expansion
of $\widetilde{\mathcal L}_{K^{\mathrm{ref}}}(\cdot,z)$ at $0$
has a quadratic remainder, uniformly for $z$ in compact subsets
of $\mathcal Z$.

\begin{proposition} \label{prop_compacity_L}
Under Assumptions~\ref{ass:model}(ii)--(iii) and
\ref{ass:reference_riccati}, the following statements hold:
\begin{itemize}
    \item[(i)]  For every compact set $\mathfrak{K}\subset \mathcal{Z}$, there exist constants $r_{\mathcal{L},\mathfrak K}>0$ and $C_{\mathcal{L},\mathfrak K}<\infty$ such that
	\begin{equation}
		B_X(0,r_{\mathcal{L}, \mathfrak K})\times\mathfrak K
		\subset
		\mathcal O_{K^{\mathrm{ref}}\times\mathcal{Z}},
		\label{eq:uniform_functional_neighborhood}
	\end{equation}
    where $\mathcal{O}_{K^{\mathrm{ref}}\times \mathcal{Z}}$ is the open set defined in Theorem \ref{thm:local_analytic_solution} and
    \begin{equation}
        \sup_{z\in \mathfrak{K}}\left\vert \widetilde{\mathcal{L}}_{K^{\mathrm{ref}%
        }}(h,z)-\widetilde{\mathcal{L}}_{K^{\mathrm{ref}}}(0,z)-\Lambda _{K^{\mathrm{ref}},z}[h]\right\vert\leq C_{\mathcal{L} ,\mathfrak{K}}\Vert
        h\Vert _{L^{1}(0,T)}^{2}  \label{eq:uniform_quadratic_funcional}
    \end{equation}
    for every $h\in B_{X}(0,r_{\mathcal{L},\mathfrak{K}})$.

    \item[(ii)] For $z\in \{0,1\}$,
    \begin{equation}
        \widetilde{\mathcal{L}}_{K^{\mathrm{ref}}}(h,z)=\widetilde{\mathcal{L}}_{K^{\mathrm{ref}}}(0,z)=D_{\eta} \widetilde{\mathcal{L}}_{K^{\mathrm{ref}}}(0,z)[h]=0
        \label{eq:quadratic_expansion_endpoints}
    \end{equation}
    for every $h\in X$ such that $(h,z)\in \mathcal{O}_{K^{\mathrm{ref}}\times \mathcal{Z}}$.
\end{itemize}
\end{proposition}

\begin{proof}
Part (i) follows by applying the argument of
Proposition~\ref{prop:compacity} to the real-analytic functional
of Proposition~\ref{prop:extended_functional_analytic},
using the derivative identification in
Theorem~\ref{thm:frechet_diff}.
For part (ii), Proposition~\ref{prop:compacity}(ii) and
$F(z,0)=0$ for $z\in\{0,1\}$ imply that
$\widetilde{\mathcal L}_{K^{\mathrm{ref}}}(\cdot,z)$
vanishes identically on its domain, so its kernel derivative
also vanishes.
\end{proof}

\subsection{Consequences for structural kernels}
\label{subsec:quadratic_expansion}

The results of Section~\ref{subsec:linearized_dual} restrict to $\mathscr K_{T}$ through %
\eqref{eq:extended_functional_coincidence}. For $S\subseteq X$ and $x\in S$,
the contingent cone is 
\begin{equation*}
T(S,x):=\left\{ v\in X:%
\begin{array}{l}
\text{there exist }t_{n}>0\text{ and }v_{n}\in X\text{ such that} \\ 
t_{n}\rightarrow 0,\quad v_{n}\rightarrow v,\quad x+t_{n}v_{n}\in S\text{
for all }n%
\end{array}%
\right\}
\end{equation*}%
(see \cite[Definition 4.1.1]{aubin1990}). All limits of kernel directions are
taken in $X=L_{\mathbb{R}}^{1}(0,T)$. Hadamard directional derivatives on $%
\mathscr K_{T}$ are understood relative to this set, as in \eqref{2909262}
(see \cite[Definition 2.2]{jimenez2009scalarization}).

\begin{corollary}\label{090926}
Suppose that Assumptions~\ref{ass:model}(ii)--(iii) hold, and let
$K^{\mathrm{ref}}\in\mathscr K_T$ and $z\in\mathcal Z$.
Then $\mathcal L_T(\cdot,z)$ is Hadamard directionally differentiable
at $K^{\mathrm{ref}}$, with
\begin{equation}\label{2909262}
\mathcal L_K'(K^{\mathrm{ref}},z)(v)
:=\lim_{\substack{t\to0^+,\ w\to v\text{ in }X\\
K^{\mathrm{ref}}+tw\in\mathscr K_T}}
\frac{\mathcal L_T(K^{\mathrm{ref}}+tw,z)
-\mathcal L_T(K^{\mathrm{ref}},z)}{t}
=\Lambda_{K^{\mathrm{ref}},z}[v]
\end{equation}
for every $v\in T(\mathscr K_T,K^{\mathrm{ref}})$.
\end{corollary}

\begin{proof}
Apply Proposition~\ref{prop_compacity_L}(i) with $\mathfrak K=\{z\}$.
By \eqref{eq:extended_functional_coincidence}, for $t>0$ and
$K^{\mathrm{ref}}+tw\in\mathscr K_T$ with
$t\|w\|_{L^1(0,T)}<r_{\mathcal L,\{z\}}$,
\[
\left|
\frac{\mathcal L_T(K^{\mathrm{ref}}+tw,z)
-\mathcal L_T(K^{\mathrm{ref}},z)}{t}
-\Lambda_{K^{\mathrm{ref}},z}[w]
\right|
\le C_{\mathcal L,\{z\}}\,t\|w\|_{L^1(0,T)}^2.
\]
As $t\to0^+$ and $w\to v$, the right-hand side tends to zero.
Continuity of $\Lambda_{K^{\mathrm{ref}},z}$ yields
\eqref{2909262}.
\end{proof}

\begin{corollary}\label{090926B}
Suppose that Assumptions~\ref{ass:model}(ii)--(iii) hold, and let
$K^{\mathrm{ref}}\in\mathscr K_T$.
\begin{itemize}
\item[(i)] For every $z\in\mathcal Z$ and every direction $v$
admissible at $K^{\mathrm{ref}}$,
\begin{equation}\label{0909262}
\mathcal L_K'(K^{\mathrm{ref}},z)(v)
=\lim_{t\to0^+}
\frac{\mathcal L_T(K^{\mathrm{ref}}+tv,z)
-\mathcal L_T(K^{\mathrm{ref}},z)}{t}
=\Lambda_{K^{\mathrm{ref}},z}[v].
\end{equation}

\item[(ii)] For every compact set $\mathfrak K\subset\mathcal Z$,
there exist constants $r_{\mathcal L,\mathfrak K},
C_{\mathcal L,\mathfrak K}>0$ such that
\begin{equation}\label{0909263}
\sup_{z\in\mathfrak K}
\left|
\mathcal L_T(J,z)-\mathcal L_T(K^{\mathrm{ref}},z)
-\Lambda_{K^{\mathrm{ref}},z}[J-K^{\mathrm{ref}}]
\right|
\le C_{\mathcal L,\mathfrak K}
\|J-K^{\mathrm{ref}}\|_{L^1(0,T)}^2
\end{equation}
for every $J\in\mathscr K_T$ with
$\|J-K^{\mathrm{ref}}\|_{L^1(0,T)}<r_{\mathcal L,\mathfrak K}$.
\end{itemize}
\end{corollary}

\begin{proof}
(i) An admissible direction belongs to
$T(\mathscr K_T,K^{\mathrm{ref}})$; taking $w=v$ in
\eqref{2909262} proves \eqref{0909262}.

(ii) Apply Proposition~\ref{prop_compacity_L}(i) with
$h=J-K^{\mathrm{ref}}$ and use
\eqref{eq:extended_functional_coincidence} at $h$ and at $0$.
\end{proof}

\begin{remark}
\label{rem:multifactor_sensitivities}
For $c,\gamma\in(0,\infty)^M$, let
$K_M(u)=\sum_{i=1}^M c_i e^{-\gamma_i u}$ and define
$\ell_M(c,\gamma;z):=\mathcal L_T(K_M,z)$.
Keep $g,\lambda,\nu,\rho$ fixed.
The parameter-to-kernel map is real-analytic into $X$ and takes
values in $\mathscr K_T$. Hence, the chain rule,
\eqref{eq:extended_functional_coincidence}, and
Theorem~\ref{thm:frechet_diff} yield, for $z\in\mathcal Z$,
\[
\begin{aligned}
\frac{\partial\ell_M}{\partial c_i}(c,\gamma;z)
&=\int_0^T B_{K_M}(u,z)e^{-\gamma_i u}\,du,\\
\frac{\partial\ell_M}{\partial\gamma_i}(c,\gamma;z)
&=-c_i\int_0^T uB_{K_M}(u,z)e^{-\gamma_i u}\,du,
\qquad i=1,\ldots,M.
\end{aligned}
\]
Thus, all $2M$ parameter derivatives use the same forward and
adjoint states for each $z$.
The kernel remainder estimate~\eqref{0909263} applies to all positive
multi-factor kernels sufficiently close to $K_M$ in $L^1(0,T)$.
\end{remark}


\section{Computational framework}
\label{sec:computational}
In this section we establish a discretization scheme for the sensitivity framework developed in Section~\ref{seccion:stability_of_CharcFunction}, based on a multi-factor approximation of the fractional kernel.

\subsection{Model and parameters}
\label{sec:num_setup}

We work with the fractional rough Heston target kernel
$K_H$ defined in~\eqref{eq:kernel}, with the values $H\in\{0.05,\,0.10,\,0.20\}$. For each $H$, with $\theta(s)\equiv\theta$, the input curve is defined by
\begin{equation}
    g_H(t)
    :=
    V_0
    +
    \lambda
    \int_0^t
    K_H(t-s)\theta\,ds
    =
    V_0
    +
    \lambda\theta
    \frac{t^{H+1/2}}{\Gamma(H+3/2)},
    \qquad
    t\in[0,T].
    \label{eq:num_input_curve_explicit}
\end{equation}

The baseline sensitivity analysis is performed at the multi-factor
reference kernel $K_{M,H}$ with the numerical functional $K\longmapsto\mathcal L_T(K,z)$ evaluated locally at $K=K_{M,H}$ while keeping $g=g_H$ fixed. In particular, $g_H$ is not recomputed under kernel perturbations. The same convention is used in the kernel-approximation robustness experiment. This is the fixed-$g$ perturbation regime adopted in Sections~\ref{sec:analytical_framework}--\ref{seccion:stability_of_CharcFunction}.

Throughout the numerical experiments, the remaining model parameters are
\begin{equation}
	\rho=-0.7,
	\qquad
	\nu=0.3,
	\qquad
	\lambda=1.5,
	\qquad
	V_0=\theta=0.04,
	\qquad
	T=1.
	\label{eq:num_params}
\end{equation}
Here, $V_0$ is the initial variance level and $\theta$ is the mean-reversion variance level. When time is measured in years, $V_0 = \theta = 0.04$ corresponds to a volatility level of $20\%$. The value $\rho=-0.7$ specifies a negative correlation between asset-price and variance shocks, consistent with the leverage effect captured by the model.

The three values $H\in\{0.05,0.10,0.20\}$ provide distinct roughness levels for assessing the numerical implementation. The first-order response ratios are reported for all three values in
Section~\ref{sec:num_locality}, whereas the temporal sensitivity profile displayed there uses $H=0.10$ as a representative case. Values below $H=0.05$ are not considered in the present numerical study.

\subsection{Multi-factor approximation of the kernel}
\label{sec:num_multifactor}

For each Hurst parameter $H$, we approximate the fractional kernel
$K_H$ of~\eqref{eq:kernel} by a finite sum of exponential modes,
following Abi~Jaber and El~Euch~\cite{abi_jaber_el_euch_2019b}:
\begin{equation}
    K_{M,H}(t)
    =
    \sum_{i=1}^M c_i^{(H)}e^{-\gamma_i^{(H)}t},
    \qquad
    c_i^{(H)}>0,
    \quad
    \gamma_i^{(H)}>0.
    \label{eq:num_Kn}
\end{equation}
When $H$ is fixed and no ambiguity can arise, we suppress the
superscript $(H)$ and write $K_M$, $c_i$, and $\gamma_i$.

The fractional kernel admits the Laplace-transform representation 
\begin{equation}
K_{H}(t)=\int_{0}^{\infty }e^{-\gamma t}\,\mu _{H}(d\gamma ),\qquad \mu
_{H}(d\gamma )=\frac{\gamma ^{-H-1/2}}{\Gamma (H+1/2)\Gamma (1/2-H)}%
\,d\gamma .  \label{eq:num_laplace_representation}
\end{equation}%
For the numerical approximation, the integration domain is truncated to
$[\gamma_{\min},\gamma_{\max}]\subset(0,\infty)$, which is partitioned as
$[\gamma_{\min},\gamma_{\max}]=\bigcup_{i=1}^{M}I_{i}$, with
$I_{i}=[x_{i-1},x_{i}]$, $i=1,\ldots,M$. The nodes are geometrically spaced,
$x_{i}=\gamma_{\min}(\gamma_{\max}/\gamma_{\min})^{i/M}$, $i=0,\ldots,M$;
for example, if $[\gamma_{\min},\gamma_{\max}]=[10^{-6},10^{6}]$, then
$x_{i}=10^{-6+12i/M}$. $\mu _{H}$ is
approximated by a finite sum of Dirac measures $\mu
_{M,H}=\sum\limits_{i=1}^{M}c_{i}\delta_{\gamma_i}$. On each cell $I_{i}$, the
continuous spectral measure is replaced by the atom $c_{i}\delta _{\gamma
_{i}}$, where 
\begin{equation*}
c_{i}=\int_{I_{i}}\mu _{H}(d\gamma ),\qquad \gamma _{i}=\frac{%
\int_{I_{i}}\gamma \,\mu _{H}(d\gamma )}{\int_{I_{i}}\mu _{H}(d\gamma )}.
\end{equation*}%
Thus, $c_{i}\delta _{\gamma _{i}}$ has the same total mass and same first
moment as $\mu _{H}|_{I_{i}}$, and contributes $c_{i}e^{-\gamma _{i}t}$ to
the Laplace transform. In numerical experiments we consider $M=20.$ Section~%
\ref{sec:num_kernel_refinement} subsequently varies both the number of
factors and the spectral truncation range.
Since $\gamma_i$ is the $\mu_H$-weighted mean of $\gamma$ over $I_i$
and $\gamma\mapsto e^{-\gamma t}$ is convex for every $t>0$, the Jensen
inequality gives
\begin{equation*}
    c_i e^{-\gamma_i t}
    \leq
    \int_{I_i}e^{-\gamma t}\,\mu_H(d\gamma).
\end{equation*}
Together with the non-negativity of the omitted spectral tails, this
yields
\begin{equation}
    K_{M,H}(t)\leq K_H(t),
    \qquad
    t>0.
    \label{eq:num_kernel_lower_bound}
\end{equation}

Consequently, the $L^1(0,T)$ approximation error can be evaluated
without sampling the singular fractional kernel at the origin:
\begin{equation}
    \|K_H-K_{M,H}\|_{L^1(0,T)}
    =
    \frac{T^{H+1/2}}{\Gamma(H+3/2)}
    -
    \sum_{i=1}^M
    c_i
    \frac{1-e^{-\gamma_iT}}{\gamma_i}.
    \label{eq:num_kernel_L1_error}
\end{equation}
We denote the corresponding relative error by
\begin{equation}
    E_K(M,H)
    :=
    \frac{
        \|K_H-K_{M,H}\|_{L^1(0,T)}
    }{
        \|K_H\|_{L^1(0,T)}
    }.
    \label{eq:num_kernel_relative_L1_error}
\end{equation}
Thus, $E_K(M,H)$ measures the integrated kernel-approximation error
relative to the total $L^1$ mass of the fractional target kernel.
For $M=20$ and $[\gamma_{\min},\gamma_{\max}]=[10^{-6},10^{6}]$,
the closed-form computation gives
\begin{equation*}
    E_K(20,0.05)=\text{4.964e-02},
    \qquad
    E_K(20,0.10)=\text{4.444e-02},
    \qquad
    E_K(20,0.20)=\text{4.179e-02}.
\end{equation*}
Each $K_{M,H}$ is a positive finite sum of exponentials and belongs to the structural multi-factor class of Remark~\ref{rem:kernel_examples}. In particular, perturbations of the existing weights of the form $h(t)= \sum\limits_{i=1}^M d_i e^{-\gamma_i t}$ are admissible at $K_M$ in the sense of Definition~\ref{def:admissible_direction}, by that remark.

\subsection{Discretization and discrete sensitivities}
\label{sec:num_riccati}

This subsection specifies the finite-resolution numerical objects used
throughout the experiments in Section~\ref{sec:num_validation}. We first discretize the multi-factor
forward Riccati equation and the continuous adjoint equation on a common
temporal grid. We then construct the numerical sensitivity kernel and
the associated quadratures, including the product-integration rule
used for the singular direction $K_H-K_{M,H}$. For the first-order validation, the resulting dual sensitivities are compared directly with finite-difference estimates over several perturbation amplitudes and temporal resolutions.

\subsubsection{Discretization of the state and adjoint equations}
\label{sec:num_forward_dual}

For a fixed Hurst parameter $H$, we suppress the $H$-dependence of $K_{M,H}$
and write $K_M$, $c_i$, and $\gamma_i$. The forward Riccati equation~%
\eqref{eq:riccati_a} and the backward dual equation~%
\eqref{eq:dual_state_backward} are discretized on the uniform grid
$t_j=j\Delta t$, $j=0,\ldots,N$, with $\Delta t=\frac{T}{N}$.

Substituting $K_{M}(t)=\sum\limits_{i=1}^{M}c_{i}e^{-\gamma _{i}t}$ into the
Volterra--Riccati equation gives
\begin{equation*}
\psi (t,z)=\sum_{i=1}^{M}c_{i}\int_{0}^{t}e^{-\gamma _{i}(t-s)}F(z,\psi
(s,z))\,ds.
\end{equation*}%
Following the multi-factor Riccati representation of Abi~Jaber and El~Euch~%
\cite{abi_jaber_el_euch_2019b}, we therefore introduce the auxiliary states
\begin{equation*}
\phi _{i}(t,z):=\int_{0}^{t}e^{-\gamma _{i}(t-s)}F(z,\psi (s,z))\,ds,\qquad
i=1,\ldots ,M,
\end{equation*}%
for which
\begin{equation*}
\psi (t,z)=\sum_{i=1}^{M}c_{i}\phi _{i}(t,z).
\end{equation*}%
Differentiating with respect to $t$, we obtain
\begin{equation*}
\partial _{t}\phi _{i}(t,z)=-\gamma _{i}\phi _{i}(t,z)+F(z,\psi
(t,z)),\qquad \phi _{i}(0,z)=0.
\end{equation*}%
We discretize this finite-dimensional system by the following semi-implicit
Euler scheme:
\begin{equation*}
\frac{\phi _{i}^{j+1}-\phi _{i}^{j}}{\Delta t}=-\gamma _{i}\phi
_{i}^{j+1}+F(z,\psi ^{j}),
\end{equation*}%
where $\phi _{i}^{j}$ and $\psi ^{j}$ approximate $\phi _{i}(t_{j},z)$ and
$\psi (t_{j},z)$, with $\phi _{i}^{0}=0$ for $i=1,\ldots ,M$ and hence
$\psi^{0}=0$. This gives the iterative scheme
\begin{equation}
\begin{array}{ll}
\phi _{i}^{0}=0, &  \\[1mm]
\phi _{i}^{j+1}=\dfrac{\phi _{i}^{j}+\Delta t\,F(z,\psi ^{j})}{1+\gamma
_{i}\Delta t}, & \qquad j=0,\ldots ,N-1,\ i=1,\ldots ,M, \\[3mm]
\psi ^{j+1}=\displaystyle\sum_{i=1}^{M}c_{i}\phi _{i}^{j+1}. &
\end{array}
\label{eq:num_riccati_recursion}
\end{equation}

For the backward equation, define $\widehat{Y}(\tau ,z):=Y_{K_{M},z}(T-\tau )$
and $\widehat{q}(\tau ,z):=q_{K_{M},z}(T-\tau )$ for $\tau \in \lbrack 0,T]$.
Setting $\tau =T-s$ in the backward equation~%
\eqref{eq:dual_state_backward} and, in the integral term, introducing the
reversed variable $r=T-t$, one obtains
\begin{equation}
\widehat{Y}(\tau ,z)-\widehat{q}(\tau ,z)\int_{0}^{\tau }K_{M}(\tau -r)%
\widehat{Y}(r,z)\,dr=\widehat{q}(\tau ,z)g_{H}(\tau ).
\label{eq:num_dual_forward}
\end{equation}%
For each exponential mode, we introduce
\begin{equation*}
\chi _{i}(\tau ,z):=\int_{0}^{\tau }e^{-\gamma _{i}(\tau -r)}\widehat{Y}%
(r,z)\,dr.
\end{equation*}%
Then
\begin{equation}
\partial _{\tau }\chi _{i}(\tau ,z)=-\gamma _{i}\chi _{i}(\tau ,z)+\widehat{Y%
}(\tau ,z),\qquad \chi _{i}(0,z)=0,  \label{eq:num_dual_auxiliary}
\end{equation}%
and~\eqref{eq:num_dual_forward} becomes
\begin{equation}
\widehat{Y}(\tau ,z)=\widehat{q}(\tau ,z)\left[ g_{H}(\tau
)+\sum_{i=1}^{M}c_{i}\chi _{i}(\tau ,z)\right] .
\label{eq:num_dual_algebraic}
\end{equation}%
With $\tau _{m}=m\Delta t$, equation~\eqref{eq:num_dual_algebraic} at
$\tau=0$ gives $\chi _{i}^{0}=0$ and $\widehat{Y}^{0}=\widehat{q}^{0}g_{H}(0)$.
Applying the same first-order semi-implicit Euler rule to
\eqref{eq:num_dual_auxiliary}, with the source $\widehat{Y}$ evaluated at
the preceding reversed-time level, gives, for $m=1,\ldots ,N$,
\begin{equation}
\chi _{i}^{m}=\frac{\chi _{i}^{m-1}+\Delta t\,\widehat{Y}^{m-1}}{1+\gamma
_{i}\Delta t},\qquad i=1,\ldots ,M,  \label{eq:num_dual_recursion_chi}
\end{equation}%
where $\chi _{i}^{m}$ and $\widehat{Y}^{m}$ approximate
$\chi_{i}(\tau _{m},z)$ and $\widehat{Y}(\tau _{m},z)$, respectively.
Following \eqref{eq:num_dual_algebraic} and \eqref{131026}, the adjoint
state is then updated algebraically as
\begin{equation}
\widehat{Y}^{m}=\widehat{q}^{m}\left[ g_{H}(\tau
_{m})+\sum_{i=1}^{M}c_{i}\chi _{i}^{m}\right] ,\qquad \widehat{q}%
^{m}=\partial _{x}F(z,\psi ^{N-m}).  \label{eq:num_dual_recursion_Y}
\end{equation}
\subsubsection{Discretization of the derivative and the sensitivity kernel}
\label{sec:num_sensitivity_quadrature}

We follow a similar approach to discretize the continuous derivative formula
established in Corollary~\ref{090926} (see also Theorem~\ref{thm:frechet_diff}).
In this sense, the derivative of the analytic extension in a direction $h$ is given by
\begin{equation}
D_{\eta}\widetilde{\mathcal{L}}_{K_M}(0,z)[h]=\Lambda_{K_M,z}[h]:=\int_{0}^{T}B_{K_{M}}(u,z)h(u)\,du,\qquad h\in X,
\label{141026}
\end{equation}%
where the sensitivity kernel is given by its analytical representation
\eqref{eq:sensitivity_kernel},
\begin{equation}
B_{K_{M}}(u,z)=\int_{0}^{T-u}F(z,\psi _{K_{M}}(\sigma ,z))Y_{K_{M},z}(\sigma
+u)\,d\sigma ,\qquad u\in \lbrack 0,T].  \label{1410262}
\end{equation}%
We assume that the lag variable $u$ is sampled on the same uniform grid as
before,
\begin{equation*}
u_{n}=t_{n}=n\Delta t,\qquad n=0,\ldots ,N,
\end{equation*}%
and, consequently, for the integration variable in \eqref{1410262} we
consider the grid
\begin{equation*}
\sigma _{j}=j\Delta t,\qquad j=0,\ldots ,N-n.
\end{equation*}%
The integrals in \eqref{141026} and \eqref{1410262} are approximated using the composite Simpson rule as implemented in
\texttt{scipy.integrate.simpson} \cite{virtanen2020scipy}. More
precisely, if we denote by $\mathrm{Simp}_{N}\left[ f\right] $ the
application of this rule on a grid of $N+1$ nodes to a generic function
$f:[0,T]\rightarrow \mathbb{C}$, the derivative \eqref{141026} is
approximated by
\begin{equation}
D_{\eta}\widetilde{\mathcal{L}}_{K_M}(0,z)[h]\approx D_{N}^{\mathrm{dual}%
}(K_{M},h,z):=\mathrm{Simp}_{N}\left[ B_{K_{M}}^{\Delta t}(\cdot
,z)\,h(\cdot )\right] ,  \label{eq:num_discrete_dual_pairing}
\end{equation}%
where $B_{K_{M}}^{\Delta t}(\cdot ,z)$ denotes the nodal values computed by
applying the same composite rule to \eqref{1410262}. At the terminal lag, we set $B_{K_M}^{\Delta t}(T,z)=0$.

For this quadrature, we assume that $h$ is pointwise defined
and its nodal values are available. The only exception is the case
$h=K_{H}-K_{M,H}$ (see Section~\ref{sec:fop}). Since $K_{H}$ is singular at
the origin, the corresponding pairing is evaluated by product integration
rather than by sampling the fractional kernel. To this end, we consider a
piecewise-linear interpolation of the sensitivity kernel. For the fractional
kernel, write $\alpha =H+\frac{1}{2}$, so that
$K_{H}(u)=\frac{u^{\alpha -1}}{\Gamma (\alpha )}$.
For the computed nodal values, we use the abbreviation
$B(u_n):=B_{K_M}^{\Delta t}(u_n,z)$. On each temporal cell
$[a,b]$, we approximate the sensitivity kernel by its linear interpolant
$B(u)\approx B(a)+\beta _{a,b}(u-a)$. The resulting integrals can then be evaluated exactly, yielding
\begin{equation*}
\int_{a}^{b}B(u)K_{H}(u)\,du\approx B(a)M_{0}(a,b)+\beta _{a,b}M_{1}(a,b),
\end{equation*}%
where $\beta _{a,b}:=\frac{B(b)-B(a)}{b-a}$ and
\begin{equation*}
\begin{array}{l}
M_{0}(a,b):=\displaystyle\int_{a}^{b}K_{H}(u)\,du=\frac{b^{\alpha }-a^{\alpha }}{\Gamma
(\alpha +1)}, \\[3mm]
M_{1}(a,b):=\displaystyle\int_{a}^{b}(u-a)K_{H}(u)\,du=\frac{1}{\Gamma (\alpha )}\left[
\frac{b^{\alpha +1}-a^{\alpha +1}}{\alpha +1}-\frac{a(b^{\alpha }-a^{\alpha
})}{\alpha }\right] .
\end{array}%
\end{equation*}%
In the same way, for the multi-factor kernel
$K_{M,H}(u)=\sum\limits_{i=1}^{M}c_{i}e^{-\gamma _{i}u}$, on each exponential mode
$e^{-\gamma u}$ we have
\begin{equation*}
\int_{a}^{b}B(u)e^{-\gamma u}\,du\approx B(a)E_{0}(a,b;\gamma )+\beta
_{a,b}E_{1}(a,b;\gamma ),
\end{equation*}%
where, by a direct computation and denoting $\ell =b-a$,
\begin{equation*}
\begin{array}{l}
E_{0}(a,b;\gamma ):=\displaystyle\int_{a}^{b}e^{-\gamma u}\,du=e^{-\gamma a}\frac{%
1-e^{-\gamma \ell }}{\gamma }, \\[3mm]
E_{1}(a,b;\gamma ):=\displaystyle\int_{a}^{b}(u-a)e^{-\gamma u}\,du=e^{-\gamma a}\frac{%
1-(1+\gamma \ell )e^{-\gamma \ell }}{\gamma ^{2}}.
\end{array}%
\end{equation*}%
Adding all the contributions, on each cell $[a,b]$ we have
\begin{equation*}
    \begin{aligned}
        \int_{a}^{b} B(u)\left(K_{H}(u)-K_{M,H}(u)\right)\,du
        &\approx B(a)M_{0}(a,b)+\beta_{a,b}M_{1}(a,b) \\
        &\quad -\sum_{i=1}^{M}c_{i}\left[
        B(a)E_{0}(a,b;\gamma_{i})
        +\beta_{a,b}E_{1}(a,b;\gamma_{i})\right],
    \end{aligned}
\end{equation*}%
and the full pairing is obtained by summing over the cells
$[t_{n},t_{n+1}]$, $n=0,\ldots ,N-1$.

\section{Numerical experiments}
\label{sec:num_validation}

Using the discretization of Section~\ref{sec:computational}, we assess the numerical implementation of the first-order sensitivity formula and study the dependence of the sensitivity on the memory lags, the concentration of the perturbation, and the multi-factor approximation of the kernel. All kernels considered in this section belong to the structural kernel class, while the reference kernels used in the numerical experiments are multi-factor kernels following the specification defined in Section \ref{sec:num_multifactor}. For each fixed Hurst parameter $H$, we keep the corresponding input curve $g_{H}$ fixed throughout each perturbation
experiment. Unless otherwise stated, we will consider $M=20$ on $[10^{-6},10^{6}]$. We compute derivatives for
specific admissible directions defined in terms of a decay rate $\gamma
_{d}>0$, such that%
\begin{equation}
h_{\gamma _{d}}(t):=c_{d}e^{-\gamma _{d}t},\qquad c_{d}:=\frac{\gamma _{d}}{%
1-e^{-\gamma _{d}T}},\qquad \Vert h_{\gamma _{d}}\Vert _{L^{1}(0,T)}=1.
\label{eq:num_direction}
\end{equation}%
For $\varepsilon \geq 0$, we consider the increment $\delta K=\varepsilon
h_{\gamma _{d}}$. Since $h_{\gamma _{d}}$ is a positive exponential mode, $%
K_{M}+\varepsilon h_{\gamma _{d}}$ remains a positive finite sum of
exponentials for every $\varepsilon \geq 0$. Hence, it belongs to the
structural multi-factor class and $h_{\gamma _{d}}$ is an admissible
direction in the sense of Definition~\ref{def:admissible_direction}.
Since $h_{\gamma_d}$ is nonnegative and has unit $L^1$-norm,
it is a probability density on $[0,T]$. We examine how this mass is distributed across lags.
For example, the fraction contained in $[0,0.4T]$ is
\begin{equation}
\int_{0}^{0.4T}h_{\gamma _{d}}(t)\,dt=\frac{1-e^{-0.4\gamma _{d}T}}{%
1-e^{-\gamma _{d}T}}.  \label{201026}
\end{equation}%
This fraction depends only on $\gamma _{d}T$ and increases with it. To
illustrate this, we consider three decay rates
\begin{equation}
\gamma _{d}\in \{0.5,\,2.0,\,8.0\},  \label{2010262}
\end{equation}%
which generate directions with progressively shorter characteristic lag
scales. A direction that is uniform on $[0,T]$ would place exactly $40\%$ of
its mass on $[0,0.4T]$. For $T=1$, the fraction is approximately $46\%$ for $%
\gamma _{d}=0.5$, $64\%$ for $\gamma _{d}=2.0$, and $96\%$ for $\gamma
_{d}=8.0$. Thus $\gamma _{d}=0.5$ spreads its mass almost uniformly over the
whole lag window, $\gamma _{d}=2.0$ moderately favors short lags, and $%
\gamma _{d}=8.0$ concentrates nearly all of its mass on short lags. We
therefore refer to $\gamma _{d}=0.5$, $2.0$ and $8.0$ as diffuse,
intermediate and short-lag concentrated, respectively.
We parametrize the Fourier contour by $v\in\mathbb R$:
\begin{equation}
z(v)=\frac{1}{2}+iv\in\mathcal Z.
\label{2010263}
\end{equation}

\subsection{Finite-difference check of the derivative formula}
\label{sec:num_linearity}

In this subsection we assess the numerical implementation
of formula~\eqref{2909262}. For a given kernel $K$, we denote by $\mathcal{L}_N(K,z)$ the numerical approximation of the Riccati functional obtained from the discrete forward solution $\{\psi_{K}^{j}\}_{j=0}^{N}$, namely
\begin{equation}
    \mathcal{L}_N(K,z)
    :=
    \operatorname{Simp}_N\left[ F(z, \psi_K^j)\,g_H(T-t_j) \right].
\end{equation}
We then compare the standard forward finite-difference quotient
\begin{equation}
D_{N,\varepsilon}^{\mathrm{FD}}(H,M,\gamma_{d},v)
:=\frac{\mathcal{L}_{N}(K_{M,H}+\varepsilon h_{\gamma_{d}},z(v))
-\mathcal{L}_{N}(K_{M,H},z(v))}{\varepsilon},\qquad \varepsilon>0,
\label{eq:num_fd_derivative}
\end{equation}
with the discrete dual pairing defined in~\eqref{eq:num_discrete_dual_pairing},
\begin{equation}
D_{N}^{\mathrm{dual}}(H,M,\gamma_{d},v)
:=\operatorname{Simp}_{N}\left[B_{K_{M,H}}^{\Delta t}(\cdot,z(v))\,
h_{\gamma_{d}}(\cdot)\right].
\label{eq:num_adjoint_prediction}
\end{equation}

We consider $H\in\{0.05,0.10,0.20\}$, $v\in\{0.5,1,2,3,5,8,12,20\}$ and $\gamma _{d}\in \{0.5,\,2.0,\,8.0\}$, and we measure the relative error
\begin{equation*}
    E_{N,\varepsilon}^{\mathrm{dual,FD}}
    :=
    \frac{\bigl|D_N^{\mathrm{dual}}-D_{N,\varepsilon}^{\mathrm{FD}}\bigr|}
         {\bigl|D_{N,\varepsilon}^{\mathrm{FD}}\bigr|}.
\end{equation*}
We compute this error for the $72$ configurations at the discretization
levels $N\in\{200,400,800\}$ over perturbation amplitudes
\[
\varepsilon\in
\{\text{1.000e-02},\,
\text{3.000e-03},\,
\text{1.000e-03},\,
\text{3.000e-04},\,
\text{1.000e-04},\,
\text{3.000e-05},\,
\text{1.000e-05}\}.
\]
Table~\ref{tab:num_dual_fd_eps} reports the average and maximum errors over the configurations.

\begin{table}[H]
    \centering
    \small
    \begin{tabular}{ccccccc}
        \toprule
        & \multicolumn{2}{c}{$N=200$}
        & \multicolumn{2}{c}{$N=400$}
        & \multicolumn{2}{c}{$N=800$} \\
        \cmidrule(lr){2-3}\cmidrule(lr){4-5}\cmidrule(lr){6-7}
        $\varepsilon$
        & $\operatorname{avg}$ & $\max$
        & $\operatorname{avg}$ & $\max$
        & $\operatorname{avg}$ & $\max$ \\
        \midrule
        1.000e-02 & 4.058e-03 & 1.306e-02 & 3.446e-03 & 5.524e-03 & 4.630e-03 & 6.840e-03 \\
        3.000e-03 & 6.005e-03 & 1.638e-02 & 2.252e-03 & 7.350e-03 & 1.047e-03 & 2.947e-03 \\
        1.000e-03 & 7.106e-03 & 1.742e-02 & 3.231e-03 & 8.381e-03 & 1.332e-03 & 3.899e-03 \\
        3.000e-04 & 7.503e-03 & 1.779e-02 & 3.626e-03 & 8.757e-03 & 1.715e-03 & 4.273e-03 \\
        1.000e-04 & 7.617e-03 & 1.789e-02 & 3.741e-03 & 8.866e-03 & 1.830e-03 & 4.383e-03 \\
        3.000e-05 & 7.657e-03 & 1.793e-02 & 3.781e-03 & 8.904e-03 & 1.870e-03 & 4.421e-03 \\
        1.000e-05 & 7.669e-03 & 1.794e-02 & 3.793e-03 & 8.915e-03 & 1.882e-03 & 4.432e-03 \\
        \bottomrule
    \end{tabular}
    \caption{Average and maximum relative errors
     of the derivative test over $72$
    configurations, for different $\varepsilon$ and $N$ ($M=20$).}
    \label{tab:num_dual_fd_eps}
\end{table}
For a fixed temporal resolution $N$, the discrepancy does not vanish as $\varepsilon$ decreases. Indeed, the finite-difference quotient converges, as $\varepsilon\to 0$, to the derivative of the discrete functional $\mathcal{L}_N$, whereas $D_{N}^{\mathrm{dual}}$ is obtained by discretizing the continuous adjoint representation of the derivative and therefore need not coincide exactly with the derivative of $\mathcal{L}_N$. Consequently, the error stabilizes at a finite-resolution level for small $\varepsilon$. This level decreases as the temporal grid is refined.

\subsection{Sensitivity across memory lags}
\label{sec:num_locality}

We study how the sensitivity kernel $B_K(\cdot,z)$, which
represents the first variation of the Riccati functional with
respect to the kernel, depends on the lag $u\in[0,T]$, from
short lags (small $u$) to the longest lags (near $T$).
We then examine how this dependence affects first-order
responses to perturbations with different lag profiles.

\subsubsection{Lag dependence of the sensitivity kernel}

In this experiment we study how strongly the Riccati functional
responds to a perturbation of the kernel at lag $u$, relative to lag $u=0$,
by computing numerically, whenever the denominator is nonzero,
the quotient
\begin{equation*}
  r(u;H,M,v):=\frac{|B_{K_{M,H}}(u,z(v))|}{|B_{K_{M,H}}(0,z(v))|}.
\end{equation*}
This ratio measures the relative memory sensitivity of the Riccati functional to the kernel at different lags $u\in[0,T]$, with respect to
$u=0$, which corresponds to the immediate past. By construction $r(0)=1$,
while $r(T)=0$ by causality: a perturbation of the kernel at lag $T$ has no
time to act before maturity, so that $B_{K_{M,H}}(T,z(v))=0$. In our
experiments $T=1$. We compute this quantity numerically for $H=0.1$ and
three values $v\in\{1,5,10\}$ of the transform variable. First, we checked
the stability of these quantities with respect to the temporal
discretization on the nested grids $N\in\{200,400,800\}$: between the
time resolution $N=400$ and $N=800$, the maximum
sup-norm change of the normalized profile is 6e-04, and the
normalizing factor $|B_{K_{M}}(0,z(v))|$ changes by at most
1e-03 in relative terms, for $v\in\{1,5,10\}$. Thus, within the numerical range considered, this ratio remains stable under temporal refinement.

\begin{figure}[htbp]
\centering
\includegraphics[width=0.53\linewidth]{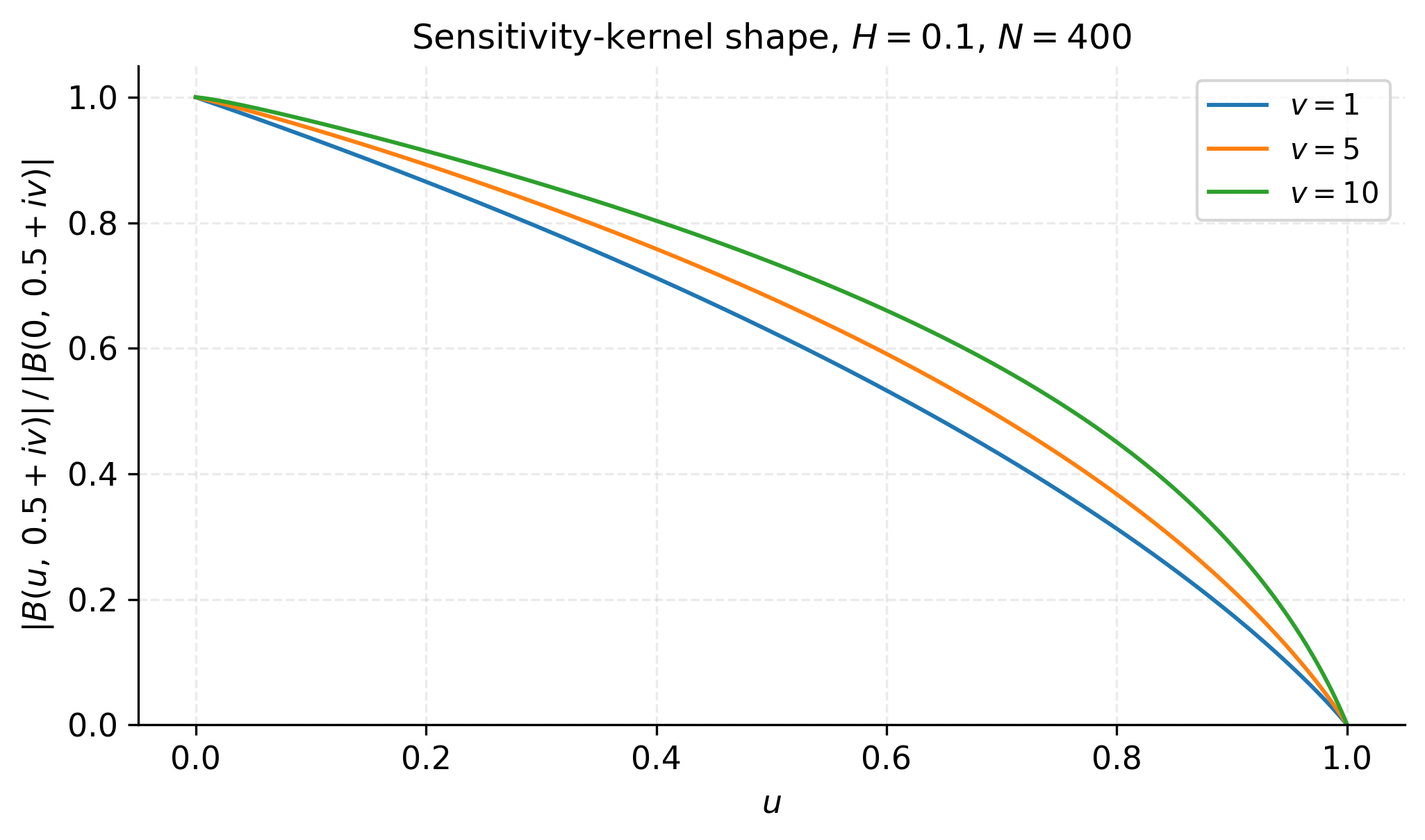}
\caption{Relative memory sensitivity of the Riccati functional $r(u;0.1,20,v)$, $v\in\{1,5,10\}$ ($N=400$).}
\label{fig:num_locality_profile}
\end{figure}

In Figure~\ref{fig:num_locality_profile}, we plot the ratio for the three values of $v$ with $N=400$.

For all three values of $v$, the computed normalized sensitivity
decreases as the lag $u$ increases, from $r(0)=1$ to $r(T)=0$.
Thus, the magnitude of the sensitivity density is larger
at shorter lags. Section~\ref{sec:lag-fo} compares the integrated
first-order responses to two specified perturbation directions.

\subsubsection{Lag dependence of the first-order response} \label{sec:lag-fo}

\begin{figure}[tbph]
\centering
\includegraphics[width=0.98\linewidth]
    {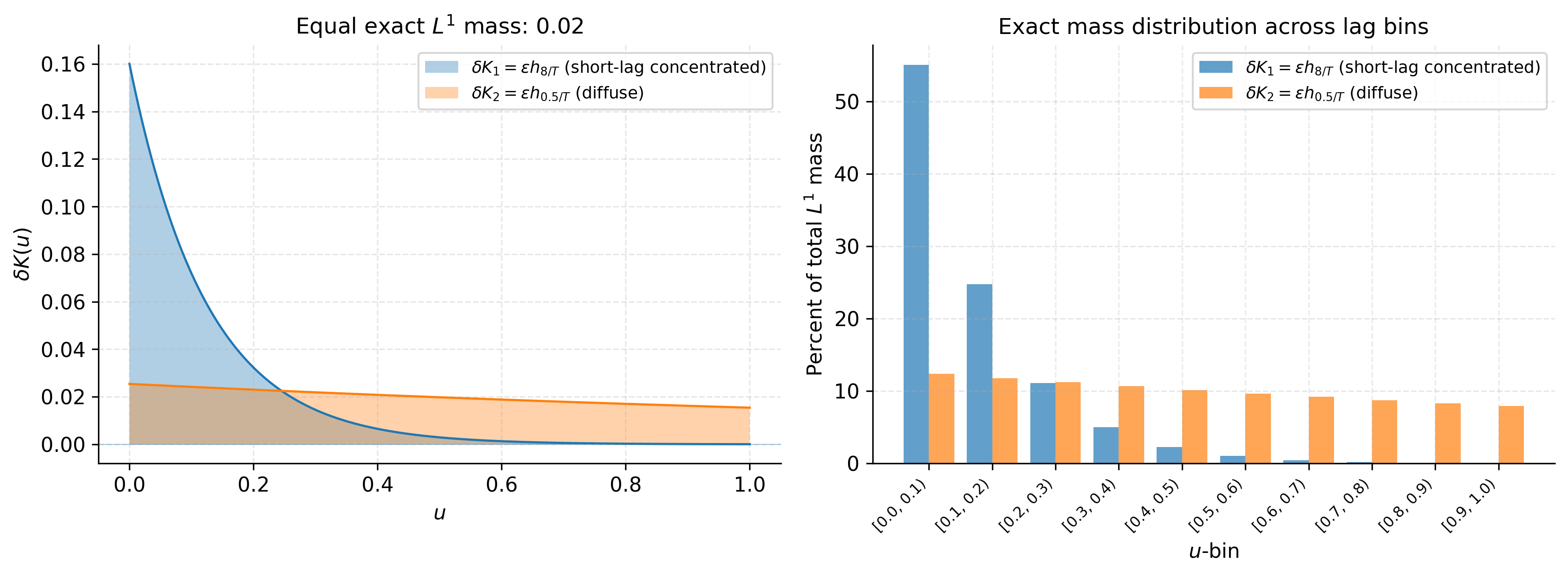}
\caption{Positive admissible increments $\protect\delta K_{1}=0.02\,h_{8/T}$
and $\protect\delta K_{2}=0.02\,h_{0.5/T}$ (left) and their mass
distributions over ten equal-width lag bins (right).}
\label{fig:num_locality_perturbations}
\end{figure}

To examine how lag allocation interacts with the sensitivity kernel and how
this affects the first-order response of the Riccati functional, we compare two
positive admissible directions with equal $L^{1}$ mass but contrasting lag
profiles, as defined in \eqref{eq:num_direction}. We consider
\begin{equation*}
h_{1}:=h_{8/T},\qquad h_{2}:=h_{0.5/T},
\end{equation*}
with $T=1$, and the corresponding admissible increments
\begin{equation}
\delta K_{\ell }:=0.02\,h_{\ell },\qquad \ell =1,2.
\label{eq:num_locality_increments}
\end{equation}
Figure~\ref{fig:num_locality_perturbations} illustrates the corresponding
increments and mass distributions over the time interval.

Using the first-variation representation of Theorem~\ref{thm:frechet_diff}
and Corollary~\ref{090926B}, for each $(H,M,v)$ with a nonzero denominator, we define the ratio of
the magnitudes of the two first-order responses, which we refer to as the
locality ratio, by
\begin{equation}
R(H,M,v):=\frac{\left\vert \int_{0}^{T}B_{K_{M,H}}(u,z(v))\delta
K_{1}(u)\,du\right\vert }{\left\vert \int_{0}^{T}B_{K_{M,H}}(u,z(v))\delta
K_{2}(u)\,du\right\vert }=\frac{\left\vert \mathcal{L}_{K}^{\prime
}(K_{M,H},z(v))[h_{1}]\right\vert }{\left\vert \mathcal{L}_{K}^{\prime
}(K_{M,H},z(v))[h_{2}]\right\vert }.  \label{eq:num_locality_ratio}
\end{equation}
$R(H,M,v)$ measures how much more strongly the Riccati functional responds to a
perturbation of the kernel concentrated at short lags than to one of the
same $L^{1}$ mass spread over all lags. Whenever the probabilistic interpretation applies, this ratio also equals the ratio of the corresponding first-order response magnitudes of the
Fourier--Laplace transform, since the common factor
$|\Phi_{K_{M,H}}(z(v))|$ cancels. We approximate this ratio
numerically for $H\in\{0.05,0.1,0.2\}$ and $v\in\{1,5,10\}$, with
$N=400$. We have checked that the ratio remains stable under refinement:
between $N=400$ and $N=800$ it changes by at most 3e-04 in
relative terms. Numerical results are displayed in
Table~\ref{tab:num_locality}.

\begin{table}[tbp]
\centering
\begin{tabular}{cccc}
\toprule
$H$ & $v=1$ & $v=5$ & $v=10$ \\
\midrule
$0.05$ & $1.467$ & $1.417$ & $1.358$ \\
$0.10$ & $1.456$ & $1.400$ & $1.334$ \\
$0.20$ & $1.437$ & $1.371$ & $1.286$ \\
\bottomrule
\end{tabular}
\caption{Numerical approximation of $R(H,20,v)$ for $h_{1}=h_{8/T}$ and $h_{2}=h_{0.5/T}$ ($N=400$).}
\label{tab:num_locality}
\end{table}

The locality ratio exceeds one in all nine tested configurations,
ranging from 1.286 to 1.467. Thus, the two directions considered
here produce different first-order response magnitudes, with the
short-lag-concentrated direction consistently producing the larger
response.

\subsection{Effect of the kernel approximation on the response} \label{sec:fop}

In this experiment, we use the first-order term to assess
the effect of approximating $K_H$ by $K_{M,H}$ on the
Riccati functional.
The term $\mathcal{L}_{K}^{\prime }(K_{M,H},z)\left(
K_{H}-K_{M,H}\right)$ is well defined, as the direction $K_{H}-K_{M,H}$ is
admissible at $K_{M,H}$: the path $(1-\varepsilon )K_{M,H}+\varepsilon
K_{H}$, $\varepsilon \in \lbrack 0,1]$, remains in the structural kernel
class covered by Assumption~\ref{ass:model}(i). Hence, by
Corollary~\ref{090926B},
\begin{equation*}
\left\vert \mathcal{L}_{K}^{\prime }(K_{M,H},z)\left(
K_{H}-K_{M,H}\right) \right\vert =\left\vert
\int_{0}^{T}B_{K_{M,H}}(u,z)\left( K_{H}(u)-K_{M,H}(u)\right)
\,du\right\vert .
\end{equation*}
Moreover, by~\eqref{eq:num_kernel_lower_bound},
$K_{H}(u)-K_{M,H}(u)\geq 0$ for $u\in (0,T]$. We therefore compute
numerically the first-order indicator
\begin{equation}
A_{\mathrm{err}}(H,M,v):=\left\vert \int_{0}^{T}B_{K_{M,H}}(u,z(v))\left(
K_{H}(u)-K_{M,H}(u)\right) \,du\right\vert .  \label{eq:num_Aerr}
\end{equation}

In our numerical experiments, we consider again $H\in \{0.05,0.1,0.2\}$ and
$v\in \{1,5,10\}$. Numerical results for $N=400$ are presented in
Table~\ref{tab:num_app}. We have previously verified that this quantity
remains stable when the temporal resolution is increased
from $N=400$ to $N=800$, with relative variations of at
most 1.7e-03. In all cases $A_{\mathrm{err}}(H,M,v)$ stays below 6e-02; it
increases with $v$ and decreases slightly as $H$ grows, so the first-order indicator is largest for large $v$ and rough kernels (small $H$).

\begin{table}[tbp]
\centering
\begin{equation*}
\begin{array}{c|ccc}
H & v=1 & v=5 & v=10 \\ \hline
0.05 & \text{7.142e-04} & \text{1.521e-02} & \text{5.737e-02} \\
0.10 & \text{6.258e-04} & \text{1.341e-02} & \text{5.085e-02} \\
0.20 & \text{5.593e-04} & \text{1.221e-02} & \text{4.710e-02}
\end{array}
\end{equation*}
\caption{Numerical values of the first-order indicator $A_{\mathrm{err}}(H,20,v)$ for $N=400$.}
\label{tab:num_app}
\end{table}

\subsection{Robustness across multi-factor approximations}
\label{sec:num_kernel_refinement}

Regarding the multi-factor approximation described in Section~\ref{sec:num_multifactor}, the preceding experiments used $M=20$ factors, with nodes geometrically spaced over the integration domain $[\gamma_{\min},\gamma_{\max}]=[10^{-6},10^{6}]$. In this section we study how the results depend on this configuration. To this end, we consider the following four specifications:
\begin{equation*}
\begin{array}{l|c|c}
\text{Label} & M & [\gamma_{\min},\gamma_{\max}] \\ \hline
\mathrm{M10} & 10 & [10^{-6},10^{6}] \\
\mathrm{M20} & 20 & [10^{-6},10^{6}] \\
\mathrm{M40} & 40 & [10^{-6},10^{6}] \\
\mathrm{M60ext} & 60 & [10^{-9},10^{9}]
\end{array}
\end{equation*}
The \textrm{M20} specification is the one used in the preceding experiments, and \textrm{M40} serves as the factor-refinement reference on the same interval. \textrm{M60ext} keeps the logarithmic cell width of \textrm{M40} but extends the interval by three decades at each end. Since its interior cells coincide with those of \textrm{M40} up to machine precision, comparing \textrm{M60ext} with \textrm{M40} isolates the effect of the spectral tails.

Given a label $s\in\{\mathrm{M10},\mathrm{M20},\mathrm{M40},\mathrm{M60ext}\}$, let $K_{s,H}$ denote the multi-factor kernel associated with specification $s$. As a first experiment, we numerically compute the relative error in the $L^{1}(0,T)$ norm between $K_{H}$ and $K_{s,H}$ for $H\in\{0.05,0.1,0.2\}$. The results are presented in Table~\ref{tab:num_mf_error}. As expected, the error decreases as $M$ increases, from order $10^{-1}$ for $\mathrm{M10}$ to order $10^{-2}$ for $\mathrm{M40}$ and $\mathrm{M60ext}$. The effect of extending the spectral interval from $\mathrm{M40}$ to $\mathrm{M60ext}$ depends on $H$; in particular, for $H=0.20$ the relative kernel error is reduced by approximately $54\%$.

\begin{table}[H]
\centering
\begin{tabular}{ccccc}
\toprule
$H$ & $\mathrm{M10}$ & $\mathrm{M20}$ & $\mathrm{M40}$ & $\mathrm{M60ext}$ \\
\midrule
$0.05$ & 1.758e-01 & 4.964e-02 & 1.423e-02 & 1.280e-02 \\
$0.10$ & 1.569e-01 & 4.444e-02 & 1.376e-02 & 1.113e-02 \\
$0.20$ & 1.257e-01 & 4.179e-02 & 2.004e-02 & 9.215e-03 \\
\bottomrule
\end{tabular}
\caption{Relative approximation error $\|K_{H}-K_{s,H}\|_{L^{1}(0,T)}/\|K_{H}\|_{L^{1}(0,T)}$.}
\label{tab:num_mf_error}
\end{table}

\begin{figure}[tbph]
\centering
\includegraphics[width=\linewidth]{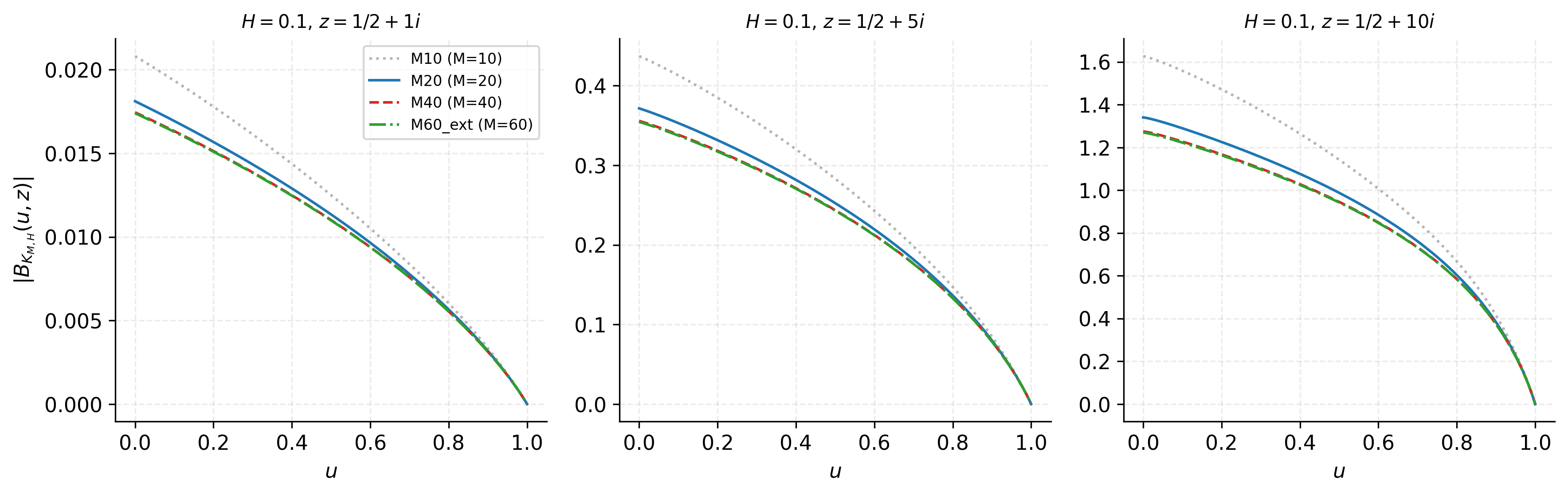}
\caption{$|B_{s}(u,\tfrac{1}{2}+iv)|$ at $H=0.10$ for $v\in\{1,5,10\}$ under the four multi-factor specifications.}
\label{fig:num_kernel_refinement_profiles}
\end{figure}

Denoting by $B_{s}(u,z):=B_{K_{s,H}}(u,z)$ the sensitivity kernel corresponding to each specification $s$, in Figure~\ref{fig:num_kernel_refinement_profiles} we plot the magnitude $\left\vert B_{s}(\cdot,z(v))\right\vert$ for $H=0.1$ and $v\in\{1,5,10\}$. Taking $\mathrm{M40}$ as the reference specification, we numerically compute the maximum nodal relative error with respect to this configuration,
\begin{equation*}
E_{B}(s):=\max\limits_{H,v}\,\max\limits_{j=0,\ldots,N-1}\frac{\left\vert B_{s}(u_{j},z(v))-B_{\mathrm{M40}}(u_{j},z(v))\right\vert }{\left\vert B_{\mathrm{M40}}(u_{j},z(v))\right\vert }.
\end{equation*}

We also compare the locality ratio and the first-order
approximation indicator studied in
Sections~\ref{sec:lag-fo} and~\ref{sec:fop}.
We denote these quantities by $R(H,s,v)$ and
$A_{\mathrm{err},s}(H,v)$, respectively, using
\eqref{eq:num_locality_ratio} and \eqref{eq:num_Aerr}
with $K_{M,H}$ replaced by $K_{s,H}$.
Their relative differences are measured by
\begin{equation*}
E_{R}(s):=\max\limits_{H,v}\frac{\left\vert R(H,s,v)-R(H,\mathrm{M40},v)\right\vert }{\left\vert R(H,\mathrm{M40},v)\right\vert },\qquad
E_{A}(s):=\max\limits_{H,v}\frac{\left\vert A_{\mathrm{err},s}(H,v)-A_{\mathrm{err},\mathrm{M40}}(H,v)\right\vert }{\left\vert A_{\mathrm{err},\mathrm{M40}}(H,v)\right\vert }.
\end{equation*}
In all cases the maxima are taken over $H\in\{0.05,0.1,0.2\}$ and $v\in\{1,5,10\}$. Numerical results for $N=400$ are presented in Table~\ref{tab:num_kernel_refinement_observables}. From the table we observe that the locality ratio $R$ remains relatively stable across the tested configurations. The sensitivity kernel shows larger variations for coarser approximations, reaching approximately $32\%$ for $\mathrm{M10}$, while $\mathrm{M60ext}$ is closest to $\mathrm{M40}$. The first-order indicator $A_{\mathrm{err}}$ is substantially more sensitive to the multi-factor specification, with the largest differences observed for $\mathrm{M10}$ and $\mathrm{M20}$.

\begin{table}[H]
\centering
\begin{tabular}{lccc}
\toprule
$s$ & $E_{B}(s)$ & $E_{R}(s)$ & $E_{A}(s)$ \\
\midrule
$\mathrm{M10}$ & 3.186e-01 & 3.202e-02 & 14.54 \\
$\mathrm{M20}$ & 5.978e-02 & 4.680e-03 & 2.709 \\
$\mathrm{M60ext}$ & 1.856e-02 & 3.680e-03 & 5.302e-01 \\
\bottomrule
\end{tabular}
\caption{Influence of the multi-factor configuration on the sensitivity kernel and related observables, with $\mathrm{M40}$ as reference ($N=400$).}
\label{tab:num_kernel_refinement_observables}
\end{table}

\section{Conclusions}
\label{sec:conclusions}

This paper develops a first-order sensitivity analysis of the rough Heston characteristic exponent with respect to its memory kernel, with the model parameters and the input variance curve held fixed. We establish the joint local real-analytic dependence of the Volterra--Riccati solution on the kernel perturbation and the transform parameter (Theorem 3.3). Analyticity carries over to the Riccati functional (Proposition 3.6), whose Fr\'{e}chet derivative admits an adjoint representation (Theorem 3.10). The derivative is given by integration against a sensitivity kernel (Lemma \ref{prop:sensitivity_kernel}), expressed in terms of an adjoint state (Definition \ref{def:dual_state}). The resulting first-order expansion has a quadratic remainder that is uniform on compact subsets of the strip (Proposition 3.11). Although the structural kernel class has empty interior in $L^{1}$, these results restrict to it: they give Hadamard directional differentiability and a uniform quadratic expansion for all sufficiently close structural kernels (Corollaries 3.12 and 3.13). The same sensitivity kernel also gives all weight and decay-rate derivatives of a multi-factor kernel (Remark 3.14).

Computationally, the theory is implemented at multi-factor reference kernels obtained by a spectral discretization of the fractional kernel. The forward Riccati and backward adjoint equations are discretized on a common grid, and the sensitivity kernel is computed by quadrature. The numerical experiments support the consistency of this implementation with the first-order theory. They also show that, in the tested regimes, the modulus of the sensitivity kernel decreases with the lag. For the two positive perturbation directions considered, the short-lag-concentrated direction produces a larger first-order response magnitude than the diffuse direction of equal $L^{1}$ mass. In addition, the sensitivity kernel provides a first-order diagnostic of the multi-factor approximation, and its qualitative robustness across multi-factor specifications is studied. From a financial modeling perspective, the representation provides a first-order assessment of how memory-kernel approximations affect the characteristic exponent used in Fourier-based derivative valuation.

Two natural questions are left for future research. First, the numerical experiments indicate that, in the tested regimes, the modulus of the sensitivity kernel decreases with the lag; identifying conditions on the kernel and on the transform parameter under which this behavior can be established analytically would clarify whether it reflects a structural property of the model. Second, it would be of interest to investigate whether the analytic and adjoint approach developed here extends to broader classes of affine Volterra models, including vector or matrix-valued Riccati--Volterra equations, and to determine which features of the present sensitivity representation persist beyond the scalar rough Heston setting.

\bigskip

\noindent\textbf{Declaration of generative AI assistance}\\[0.5em]
During manuscript preparation, ChatGPT (OpenAI, GPT-5.6) was used for language editing and wording refinement, organizational suggestions, checks of notational and textual consistency, and assistance with code review. The mathematical arguments, numerical results, code, references, and final manuscript content were independently reviewed and verified by the authors, who take full responsibility for the work.

\bigskip

\noindent\textbf{Data availability statement}\\[0.5em]
The numerical results reported in this article were generated computationally from the model and parameter specifications described in the manuscript. No external datasets were used.

\bigskip

\noindent\textbf{Conflict of interest statement}\\[0.5em]
The authors declare that they have no conflict of interest.

\bigskip

\noindent\textbf{Acknowledgments}\\[0.5em]
This research has been partially supported by MICIU / AEI / 10.13039 / 501100011033 / and by ERDF, EU (Projects PID2024-156273NA-I00 (Sama) and PID2024-158823NB-I00 (Escudero)), by MICIU / AEI / 10.13039 / 501100011033 / European Union NextGenerationEU / PRTR (Project CPP2024-011557 (Escudero and Sama)), and by COST (European Cooperation in Science and Technology) through COST Action Stochastica CA24104 (Escudero).

\appendix
\setcounter{equation}{0}
\setcounter{table}{0}
\setcounter{figure}{0}
\renewcommand{\theequation}{A.\arabic{equation}}
\renewcommand{\thetable}{A.\arabic{table}}
\renewcommand{\thefigure}{A.\arabic{figure}}

\addcontentsline{toc}{section}{References}
\bibliographystyle{acm} 
\bibliography{kernel_sensitivity_v2.bib} 

\clearpage

\noindent
Carlos Escudero\\
Departamento de Matem\'aticas Fundamentales\\
Facultad de Ciencias\\
Universidad Nacional de Educaci\'on a Distancia\\
\texttt{cescudero@mat.uned.es}

\bigskip

\noindent
Francisco J. Luguera-Par\'uas\\
Departamento de Matem\'aticas Fundamentales\\
Facultad de Ciencias\\
Universidad Nacional de Educaci\'on a Distancia\\
\texttt{fluguera1@alumno.uned.es}

\bigskip

\noindent
Miguel Sama\\
Departamento de Matem\'atica Aplicada\\
Escuela T\'ecnica Superior de Ingenieros Industriales\\
Universidad Nacional de Educaci\'on a Distancia\\
\texttt{msama@ind.uned.es}


\end{document}